\documentclass{aa}
\usepackage{graphicx}
\usepackage{bm}
\usepackage{times}
\usepackage{amsmath}
\usepackage{amssymb}
\usepackage{natbib}
\usepackage{color}
\usepackage{hyperref}
\bibpunct{(}{)}{;}{a}{}{,}
\graphicspath{{./fig/}{./png/}}
\newcommand{\uuu}{{\bm u}}

\newcommand{\BBB}{{\bm B}}

\newcommand{\Peff}{\mathcal{P}_{\rm eff}}
\newcommand{\EQ}{\begin{equation}}
\newcommand{\EE}{\end{equation}}
\newcommand{\EQA}{\begin{eqnarray}}
\newcommand{\EEA}{\end{eqnarray}}

\newcommand{\pd}{\partial}

\newcommand{\DIV}{\vec{\nabla} \cdot }

\newcommand{\mean}[1]{\overline{#1}}
\newcommand{\meanv}[1]{\overline{\bm #1}}
\newcommand{\cP}{c_{\rm P}}
\newcommand{\cs}{c_{\rm s}}

\newcommand{\urms}{u_{\rm rms}}

\newcommand{\Beq}{B_{\rm eq}}

\newcommand{\Hp}{H_{p}}
\newcommand{\Hrho}{H_{\rho}}

\newcommand{\chiSGS}{\chi_{\rm SGS}}
\newcommand{\chiSGSm}{\mean\chi_{\rm SGS}}

\newcommand{\Pm}{{\rm Pm}}
\newcommand{\Rem}{{\rm Rm}}
\newcommand{\Pra}{{\rm Pr}}
\newcommand{\PraSGS}{{\rm Pr}_{\rm SGS}}
\newcommand{\Ra}{{\rm Ra}}

\newcommand{\Rey}{{\rm Re}}

{}
\def\onethird{{\textstyle{1\over3}}}
\def\onehalf{{\textstyle{1\over2}}}

\newcommand{\Fig}[1]{Fig.~\ref{#1}}

\newcommand{\G}{\,{\rm G}}

\newcommand{\const}{{\rm const}  {}}

\begin{document}

\authorrunning{K\"apyl\"a et al.}
\titlerunning{Magnetic flux concentrations from turbulent stratified convection}

   \title{Magnetic flux concentrations from turbulent\\ stratified convection}

   \author{P. J. K\"apyl\"a
          \inst{1,2,3}
          \and
          A. Brandenburg
          \inst{3,4,5,6}
           \and
          N. Kleeorin
          \inst{7,3}
           \and
          M. J. K\"apyl\"a
          \inst{1}
           \and
          I. Rogachevskii
          \inst{7,3}
          }
   \offprints{\email{petri.kapyla@aalto.fi}
              }

   \institute{ReSoLVE Centre of Excellence, Department of Computer Science, Aalto University, PO Box 15400, FI-00076 Aalto, Finland
         \and Department of Physics, Gustaf H\"allstr\"omin katu 2a
              (PO Box 64), FI-00014 University of Helsinki, Finland
         \and NORDITA, KTH Royal Institute of Technology and Stockholm University, Roslagstullsbacken 23, SE-10691 Stockholm, Sweden
         \and Department of Astronomy, AlbaNova University Center,
              Stockholm University, SE-10691 Stockholm, Sweden
         \and JILA and Department of Astrophysical and Planetary Sciences,
              Box 440, University of Colorado, Boulder, CO 80303, USA
         \and Laboratory for Atmospheric and Space Physics,
              3665 Discovery Drive, Boulder, CO 80303, USA
         \and Department of Mechanical Engineering, Ben-Gurion University
of the Negev, PO Box 653, Beer-Sheva 84105, Israel
}

   %\date{Received ? / Accepted ?}
\date{\today,~ $ $Revision: 1.230 $ $}

   \abstract{
     The mechanisms that cause the formation of sunspots
     are still unclear.
   }{%
     We study the self-organisation of initially uniform
       sub-equipartition magnetic fields
       by highly stratified turbulent convection.
   }%
   {   We perform simulations of magnetoconvection in
       Cartesian domains representing the uppermost $8.5$--$24$~Mm of the
       solar convection zone with the horizontal size of the domain
       varying between 24 and 96~Mm.
     The density contrast in the 24~Mm deep models is more than
     $3\times10^3$ or eight density scale heights, corresponding to
     a little over 12 pressure scale heights.
     We impose either a vertical or a horizontal uniform magnetic field in
     a convection-driven turbulent flow in setups where no
     small-scale dynamos are present. In the most highly stratified
     cases we employ the reduced sound speed method to relax the time
     step constraint arising from the high sound speed in the deep
     layers.
     We model radiation via the diffusion approximation and
     neglect detailed radiative transfer in order to concentrate on
     purely magnetohydrodynamic effects.
   }%
   {
     We find that super-equipartition magnetic flux concentrations are
     formed near the surface in cases with moderate and high density
     stratification, corresponding to domain depths of $12.5$ and
     $24$~Mm. The size of the concentrations increases as the box size increases and
     the largest structures (20~Mm horizontally near the surface) are obtained in
     the 24~Mm deep models.
     The field strength in the concentrations is in the range of
     $3$--$5$~kG, almost independent of the magnitude of the imposed
     field.
     The concentrations grow approximately linearly in time.
     The effective magnetic pressure measured in the simulations is positive near
     the surface and negative in the bulk of the convection zone.
     Its derivative with
     respect to the mean magnetic field, however, is positive in
     the majority of the domain, which is unfavourable for the
     negative effective magnetic pressure instability (NEMPI) to operate.
     Simulations in which a passive vector field is evolved do not show a
     noticeable difference from magnetohydrodynamic runs in terms of
     the growth of the structures.
     Furthermore, we find that magnetic
     flux is concentrated in regions of converging flow corresponding
     to large-scale supergranulation convection pattern.
   }%
   {
     The linear growth of large-scale flux concentrations implies that
     their dominant formation process
     is tangling of the large-scale field
     rather than an instability.
     One plausible mechanism explaining both the linear growth and the
     concentration of the flux in the regions of converging
     flow pattern is flux expulsion.
     A possible reason for the absence of NEMPI is the
     fact that the derivative of the effective magnetic pressure with
     respect to the mean magnetic field has an unfavourable sign.
     Furthermore, there may not be sufficient scale separation,
     which is required for NEMPI to work.
   }%

   \keywords{   convection --
                turbulence --
                sunspots
               }

  \maketitle

%____________________________________________________________

\section{Introduction}

The current paradigm of sunspot formation relies on the
existence of strong magnetic flux tubes (of the order of $10^5\G$)
created by some unknown mechanism
at the base of the convection zone or just below it. Their buoyant rise to the solar
surface is thought to lead to sunspot formation \citep{Pa55a}. This
idea has also profoundly influenced solar dynamo modeling: in the
so-called flux transport models a highly non-local $\alpha$-effect is used to
parametrise the rise of toroidal flux tubes from the tachocline to
form poloidal fields near the surface. A single cell meridional flow
is then supposed to carry the surface poloidal field back to the
tachocline where it is sheared back to toroidal form and amplified to close the dynamo loop
\citep[e.g.][]{CSD95,DC99,DG06,CCJ07}.

Although superficially plausible, these concepts face several
theoretical difficulties: the generation and storage of sufficiently
strong magnetic fields has proven to be difficult
\citep[e.g.][]{GCS10,GK11}, the stability of the tachocline
has been questioned in the case of such strong fields \citep{ASR05},
and there are helioseismic indications \citep{STR13,ZBKDH13} and
numerical evidence \citep[e.g.][]{KKB14,PCM15,FM15} that the meridional
circulation pattern of the Sun is likely to consist of multiple cells.
Lastly, the rotational speeds of active regions
are also consistent with
the idea that spots are formed near the surface \citep{Br05} which
calls for a new mechanism of sunspot formation.

One possibility is the negative effective magnetic pressure
instability (NEMPI) in strongly stratified turbulence,
which results from the fact that
large-scale magnetic field causes a reduction of the total
(hydrodynamic plus magnetic) turbulent pressure. As a result,
the effective magnetic pressure (the sum of non-turbulent and turbulent
contributions to the large-scale magnetic pressure)
becomes negative and a large-scale MHD instability
can become excited.
This instability does not produce new magnetic flux,
but redistributes the large-scale magnetic field so that
the regions with super-equipartition magnetic fields
are separated by regions with weak magnetic field.
This effect has been thoroughly studied analytically
\citep[e.g.][]{KRR89,KRR90,KMR93,KMR96,KR94,RK07} and more recently numerically
\citep[e.g.][and
references therein]{BKR10,BKKR12,KBKR12,KBKMR12}. Further numerical studies
have confirmed the existence of NEMPI in direct numerical simulations (DNS)
of forced turbulence with weak imposed horizontal \citep{BKKMR11}
and vertical \citep{BKR13} magnetic fields, and in a two-layer
system with an upper unforced coronal layer and a lower
forced layer \citep{WLBKR13,WLBKR15}.
With NEMPI, even uniform, sub-equipartition,
magnetic fields can lead to flux concentrations
if there is sufficient scale separation between the
forcing scale and the size of the domain
in strongly stratified turbulence.
This mechanism is compatible with
a shallow origin of sunspots. Furthermore, numerical simulations of
convective dynamos produce diffuse magnetic fields throughout the
convection zone \citep[e.g.][]{GCS10,KMB12,YGCR15, ABMT15}, which
could act as the seed field for NEMPI.

An entirely different kinematic process to form magnetic concentrations is flux
expulsion where magnetic field is expelled from regions of rapid
motion. A classical example is a convection cell where fields are
swept away from the diverging upflows of granules to intergranular
lanes and vertices to form concentrations \citep{W66}. Results from
relatively weakly stratified numerical simulations of convection can
be explained by this process \citep[e.g.][]{TWBP98,KKWM10,TP13} but
its role in the presence of strong stratification has not previously
been studied. A further possibility is a mean-field instability caused by the
suppression of turbulent heat flux by magnetic fields.
Such a suppression causes a concentration of the magnetic field
which causes enhanced quenching of convection
and further concentration of the field \citep{KM00}.

Realistic numerical simulations of solar surface convection in
Cartesian domains including
radiation transport and ionization are now routinely used to study
the structure of sunspots and active regions
\citep[e.g.][]{RSCK09,RSK09,CRTS10}. These
models, however, do not address the question of sunspot formation, as
the field configuration is controlled by prescribed boundary
conditions at the base of the layer. A more self-consistent
approach is adopted in the model of \cite{SN12} where a
1~kG purely horizontal field is advected through the bottom boundary
of the highly stratified gas in their domain, mimicking the emergence
of flux from deeper layers. In this setup, encompassing the top 20~Mm
of the solar convection zone, the magnetic field ultimately forms a
magnetic structure which is buoyantly unstable and rises to the surface
to form a small bipolar spot pair. The authors
relate the formation of the structure with the large-scale supergranular
convection in the deep layers of their simulation, which would be
qualitatively consistent with flux expulsion. However, this
conclusion is based on a single experiment and these results have yet
to be put into a theoretical framework, that would allow these
results to be generalised to other conditions.

Based on the recent success in the detection of NEMPI in forced
turbulence setups, it is of great interest to study whether it can also
be excited in convection, especially in circumstances similar to those
in the study of \cite{SN12}. Earlier work on the subject did reveal
the existence of a negative effective magnetic pressure caused by
a negative contribution of turbulent convection,
but NEMPI was not observed \citep{KBKMR12,KBKMR13}.
The failure to excite NEMPI in the earlier models is possibly related to too low
density stratification and poor separation of scales. We set out to study
magnetic structure formation with improved high-resolution local
convection simulations that are constructed so that they should be
more favourable for NEMPI to be excited. However, we also consider
other processes, namely flux expulsion, that can explain magnetic
structure formation in our simulations.

\section{The Model} \label{sect:model}

As a basis for our model we use the setup from \cite{KBKMR13} with
several improvements in order to increase the density stratification
and scale separation. Firstly, we use a thin cooling layer at the top
where the temperature is cooled toward a constant value. As a
consequence, the density decreases exponentially in this region. Secondly,
instead of regular constant kinematic viscosity, we apply a
version of Smagorinsky viscosity \citep{HB06} in the highest resolution cases to
increase the effective fluid Reynolds number
and degree of scale separation. Thirdly, to facilitate computations with the increased
stratification, which leads to low Mach numbers at the base of the
convectively unstable layer, we apply the so-called reduced sound
speed method \citep{Re05,HRYIF12,HRY14} to alleviate the time step
constraint.

We solve the compressible hydromagnetics equations,
\begin{equation}
\frac{\pd \bm A}{\pd t} = {\bm u}\times{\bm B} - \eta \mu_0 {\bm J},
\end{equation}
\begin{equation}
\frac{\pd \rho}{\pd t} = -\frac{1}{\xi^2} \bm\nabla\cdot(\rho \bm{u}),
\end{equation}
\begin{equation}
\frac{D\bm{u}}{Dt} = \bm{g} + \frac{1}{\rho}
\left[\bm\nabla \cdot(2\nu\rho\bm{\mathsf{S}})-\bm\nabla p + {\bm J} \times {\bm B}\right],
\end{equation}
\begin{equation}
T\frac{D s}{Dt} = \frac{1}{\rho}\left[\bm\nabla \cdot (K \bm\nabla T + \chiSGS \rho T \bm\nabla s) + \mu_0\eta {\bm J}^2\right] + 2\nu \bm{\mathsf{S}}^2 + \Gamma ,
\label{equ:ss}
\end{equation}
where ${\bm A}$ is the magnetic vector potential, $\bm{u}$ is the
velocity,
${\bm B} ={\bm B}_0 + \bm\nabla\times{\bm A}$ is the magnetic field,
${\bm B}_0$ is the imposed magnetic field,
${\bm J} =\mu_0^{-1}\bm\nabla\times{\bm B}$ is the current density,
$\eta$ is the magnetic diffusivity, $\mu_0$ is the vacuum
permeability, $\rho$ is the density, $\xi$ is the sound speed
reduction factor, $D/Dt = \pd/\pd t + \bm{u} \cdot \bm\nabla$ is the
advective time derivative, ${\bm g}=-g\hat{\bm e}_z=\const$ is the
gravitational acceleration, $\nu$ is the kinematic viscosity, $K$ is the radiative heat
conductivity, $\chiSGS$ is the subgrid scale (SGS) heat
conductivity, $\Gamma$ describes the cooling applied at the surface,
$s$ is the specific entropy, $T$ is the temperature, and $p$ is the
pressure. The fluid obeys the ideal gas law with $p=(\gamma-1)\rho e$,
where $\gamma=c_{\rm P}/c_{\rm V}=5/3$ is the ratio of specific heats
at constant pressure and volume, respectively, and $e=c_{\rm V} T$ is
the internal energy. The traceless rate-of-strain tensor
$\mbox{\boldmath ${\sf S}$}$ is given by
\begin{equation}
{\sf S}_{ij} = \onehalf (U_{i,j}+U_{j,i}) - \onethird \delta_{ij} \DIV \bm{U}.
\end{equation}
For the viscosity we either apply constant kinematic viscosity
$\nu=\nu_0$ or the Smagorinsky viscosity
$\nu=(C_k\Delta)^2\sqrt{\mbox{\boldmath ${\sf S}$}^2}$, where $\Delta$
is the filtering scale which is here the grid spacing, and
$C_k=0.35$ has been found suitable.

For the sound speed reduction factor $\xi$ we either use a constant
value of unity, when there is no reduction, or a profile that matches the
vertical stratification of sound speed. The latter choice leads to an
effective sound speed which is constant in the whole domain. In the
latter case the gain in the time step is roughly a factor of five in
comparison to the $\xi=1$ case in the runs with the greatest vertical
extent.

The depth of the layer is $L_z=d$ and the horizontal extents in the $x$ and
$y$ directions are $L_{\rm h}=4\,d$. We consider three values of $L_z$ that
correspond to $8.5$, $12.5$, and $24$~Mm in physical units, see
Sect.~\ref{sec:solpar}.
The top and bottom boundaries are impenetrable and stress free for the
flow
\begin{eqnarray}
\frac{\pd u_x}{\pd z} &=& \frac{\pd u_y}{\pd z} = u_z =0,
\end{eqnarray}
and the magnetic field (not including the imposed field) is assumed to
be either perfectly vertical or horizontal field:
\begin{eqnarray}
B_x &=& B_y = 0 \hspace{2.5cm} \mbox{(vertical field)}, \\
\frac{\pd B_x}{\pd z} &=& \frac{\pd B_y}{\pd z} = B_z =0
\hspace{1cm} \mbox{(perfect conductor)}.
\end{eqnarray}
The energy flux at the lower boundary is fixed
\begin{eqnarray}
F_{\rm bot} = - K\frac{\pd T}{\pd z} - \chiSGS \rho T \frac{\pd s}{\pd z}.
\end{eqnarray}
At the top boundary the temperature is fixed.
The radiative conductivity is given by $K=\rho c_{\rm P}\chi$,
where $\chi$ is assumed constant throughout the domain. For
$\chiSGS$ we use a profile so that it has a constant value
$0.1\chiSGSm$ in the lower 20 per cent of the domain and connects
smoothly to a value $\chiSGSm$ in the middle part. In the layer
consisting of the uppermost four per cent of the box $\chiSGS$ drops
smoothly to zero.

To maximise the density contrast within the convection zone,
we omit a stably stratified layer below that one. We add a nearly
isothermal cooling layer at the top where the density stratification
is also strong.
The cooling term $\Gamma$ relaxes the temperature toward the value at
the surface
\begin{eqnarray}
\Gamma = f(z) L_0 \frac{T-T_{\rm cool}}{T_{\rm cool}},
\end{eqnarray}
where $f(z)=1$ in the cooling layer above $z=z_{\rm cool}$ and zero
elsewhere, and $L_0$ is a cooling luminosity.
The pressure scale height in the cooling layer is given by
\begin{eqnarray}
H_{\rm p}^{(\rm cool)} = \frac{c_{\rm V} (\gamma -1) T_{\rm cool}}{gd}.
\end{eqnarray}

\begin{figure}[t]
\centering
\includegraphics[width=\columnwidth]{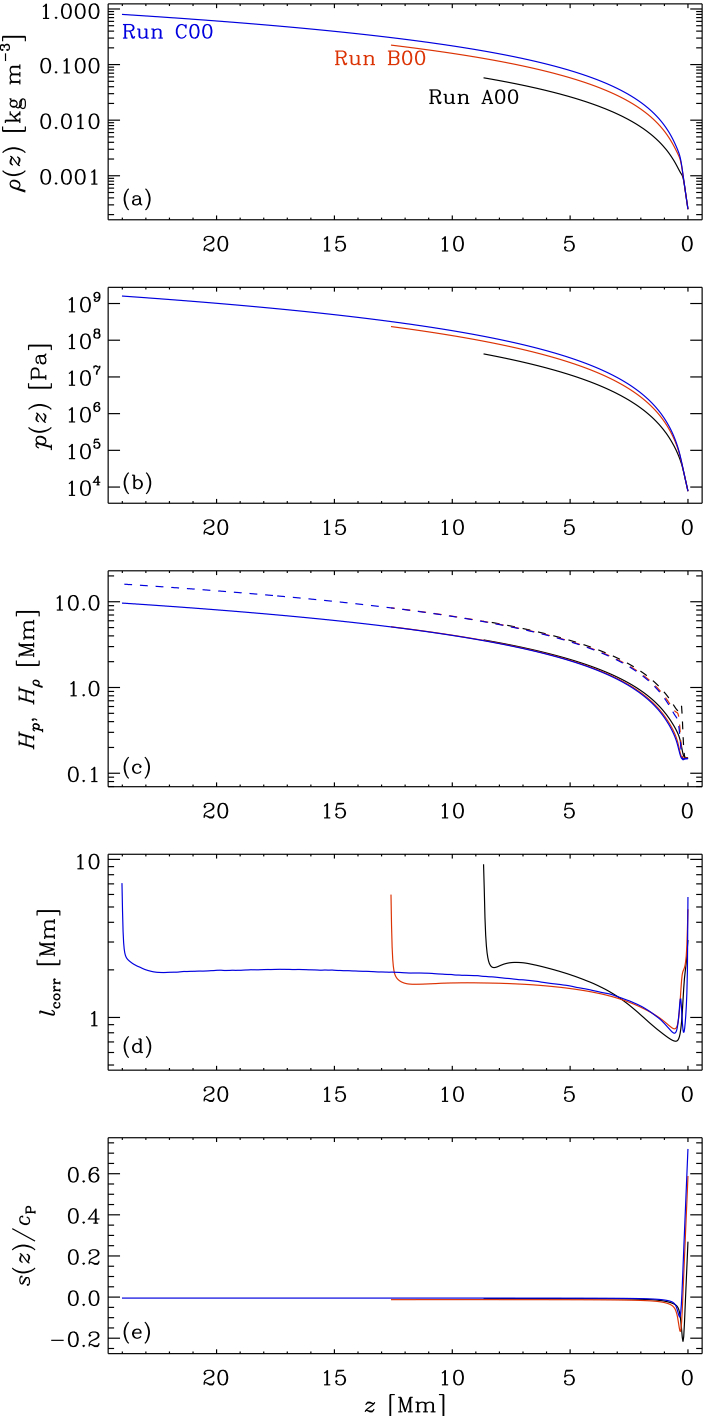}
\caption{Comparison of the stratifications of our three hydrodynamic runs
A00 (black), B00 (red), and C00 (blue) showing density (a), pressure (b),
the density (solid lines) and pressure scale heights (dashed lines) (c),
correlation length $l_{\rm corr}=2\pi/k_\omega$ (d), and specific
entropy (e).}
\label{fig:pprofs}
\end{figure}

In this setup convection transports the majority of the flux whereas
radiative diffusion is only important near the bottom of the
domain. We start hydrodynamic progenitor runs from isentropic
stratifications throughout and apply the cooling above $z_{\rm cool}$.
In the thermally relaxed states we obtain density contrasts,
$\Gamma_\rho=\rho_{\rm bot}/\rho_{\rm top}$,
of $230$ (Set~A), $900$ (Set~B), and $3.2\cdot10^3$ (Set~C) in
the three sets of runs, see Table~\ref{tab:sets}.
The corresponding density contrasts within the convectively
unstable region are denoted as $\Gamma_\rho^{\rm conv}$, which vary
between 60--320 from Set A--C, respectively.
The horizontally averaged profiles of density and
pressure along with the corresponding scale heights, and the specific
entropy, are shown in Fig.~\ref{fig:pprofs}.

\subsection{Diagnostics}

We define the fluid and magnetic Reynolds numbers as
\begin{eqnarray}
\Rey = \frac{\urms}{\nu k_1}, \hspace{0.5cm} \Rem = \frac{\urms}{\eta k_1},
\end{eqnarray}
where $\urms$ is the rms value of the volume averaged velocity and
$k_1=2\pi/d$. We also define Prandtl numbers as
\begin{eqnarray}
\Pra = \frac{\nu}{\chi}, \hspace{0.5cm} \PraSGS = \frac{\nu}{\chi_{\rm SGS}},
\hspace{0.5cm} \Pm = \frac{\nu}{\eta},
\end{eqnarray}
and the Rayleigh number
\begin{eqnarray}
\Ra = \frac{g d^4}{\nu \chiSGS} \left(-\frac{1}{\cP}
\frac{{\rm d}s}{{\rm d}z} \right)_{z_{\rm m}},
\end{eqnarray}
where $z_{\rm m}=0.5\,d$ denotes the middle of the unstable layer.
In many of the simulations considered here, only the magnetic
Reynolds number is
well defined because we are using the Smagorinsky scheme for
the viscosity.
The normalised energy flux is given by
\begin{eqnarray}
\mathcal{F} = \frac{F_0}{(\rho \cs^3)_{\rm bot}},
\end{eqnarray}
where the input flux $F_0$, density $\rho$, and the sound speed
$\cs=\sqrt{\gamma p/\rho}$ are evaluated at the lower boundary.
We also define the Taylor microscale wavenumber
\begin{eqnarray}
k_\omega=\frac{\omega_{\rm rms}}{\urms},
\label{equ:komega}
\end{eqnarray}
which is used in the estimate of the correlation length $l_{\rm
  corr}=2\pi/k_\omega$ plotted in Fig.~\ref{fig:pprofs}(d).
Here $\bm\omega=\bm\nabla\times{\bm u}$.
In isotropically forced turbulence, $k_\omega$ is proportional to the
square root of the Reynolds number based on the integral wavenumber;
see Fig.~3 of \cite{CB13}.
Calculating the integral wavenumber is usually done via energy spectra,
but in stratified convection those spectra change significantly with
height, making this approach less practical.
The equipartition field strength is defined as
\begin{eqnarray}
  \Beq(z)=\left\langle\mu_0 \rho \uuu^2\right\rangle_{xy}^{1/2}.
\end{eqnarray}
In the following, averaging over the $xy$ plane is
also indicated by an overbar. We typically apply concurrent horizontal
and temporal averages to present our results. However, in the cases
with an imposed horizontal field we sometimes average along the
imposed field which is mentioned explicitly when applied. In order to
extract the large-scale flows generated in the simulations we perform
temporal averaging over snapshots without spatial averaging in
Sect.~\ref{sec:peff}.
We use grid resolutions of up to $1024^3$.
The computations were performed with the {\sc Pencil
  Code}\footnote{https://github.com/pencil-code/}.

\subsection{Modeling strategy}

Making the simulation domain deeper and thus increasing the density
stratification in convection simulations
implies that the sound speed in the deep layers becomes very large and
limits the time step. We use the aforementioned reduced sound speed method to
overcome this problem.
Furthermore, the pressure scale height near the surface becomes small,
necessitating high spatial resolution.
We also choose the input flux sufficiently low such
that the Mach number near the surface remains sufficiently below unity.
This implies a small radiative diffusivity $\chi=K/\rho c_{\rm P}$ and
a long thermal relaxation time, which would require prohibitive computational
resources if the simulations were run from scratch.

To address the aforementioned difficulties, we first
evolve hydrodynamic runs where the horizontal extent is reduced by a
factor between four and eight to save computational time. Once these
runs have relaxed sufficiently, we replicate them onto a larger horizontal
domain and introduce a localised small-scale perturbation in one of
the subdomains to break the symmetry introduced in the
replication. The system loses the symmetry within a few convective
turnovers. We continue to run these hydrodynamical progenitor runs
for several tens of convective turnover times before introducing a
uniform magnetic field into the system.

\begin{figure}[t]
\centering
\includegraphics[width=\columnwidth]{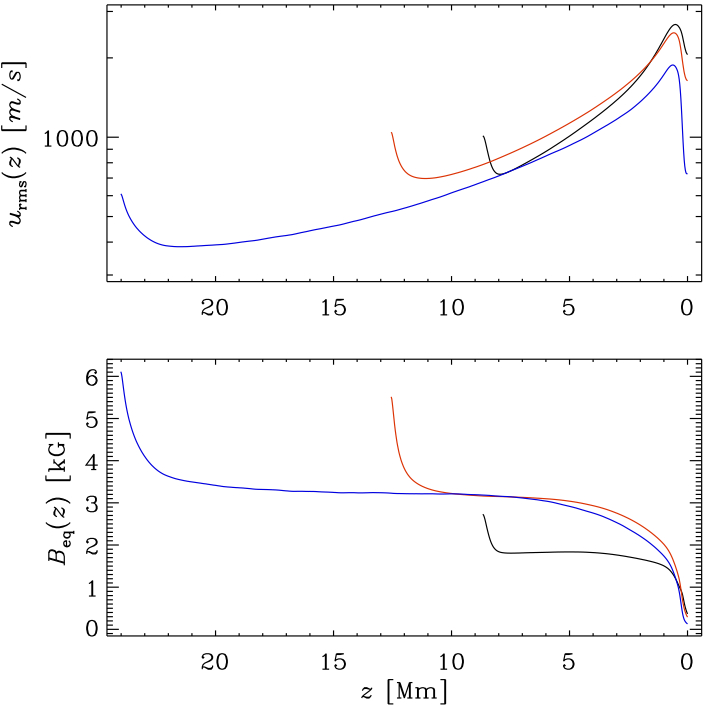}
\caption{Profiles of horizontally averaged rms velocity $\urms$ (a)
  and equipartition magnetic field $\Beq$ (b) from the same runs as in
  Fig.~\ref{fig:pprofs} in units of m~s$^{-1}$ and kG, respectively.
  }
\label{fig:purms}
\end{figure}

\begin{table}[t!]
\centering
\caption[]{Summary of the sets of runs.}
       \label{tab:sets}
%       \vspace{-0.5cm}
      $$
          \begin{array}{p{0.05\linewidth}cccrcc}
            \hline
            \hline
            \noalign{\smallskip}
Set & \mbox{Grid} & L_{\rm h}^2\times L_z [{\rm Mm}]  & L_{\rm c} [{\rm Mm}] & \Gamma_\rho & ~\Gamma_\rho^{\rm conv}~~  & \mathcal{F} [10^{-6}] \\
\hline
A   & 576^2\times288 & 34^2 \times 8.5  & 0.17 &  230 &~~60 & 7.0 \\
B   & 512^3          &~~50^2 \times 12.5 & 0.25 &  900 & 110 & 1.7  \\
C   & 1024^3         & 96^2 \times 24~   & 0.36 & 3200 & 320 & 0.10  \\
\hline
          \end{array}
          $$
\tablefoot{Here $L_{\rm c}$ is the depth of the cooling layer. In
  Set~A $\Ra=1.2\cdot10^8$, $\PraSGS=1$, and $\Pra=10$. Runs in Sets~B
  and C employ Smagorinsky viscosity and the reduced sound speed method.
In those two sets, $\urms/\nu_{\rm rms} k_1$
is around 480 and 1200, respectively.
}\end{table}

\begin{table}[t!]
\centering
\caption[]{Summary of the runs.}
%The runs are found in the folder:
%\texttt{petri/convection/nempi}
       \label{tab:runs}
%       \vspace{-0.5cm}
      $$
          \begin{array}{p{0.065\linewidth}cccccccccc}
            \hline
            \hline
            \noalign{\smallskip}
Run & \Rey & \Rem & B_{\rm rms} & \BBB_0\hat{\bm e}_y & \BBB_0\hat{\bm e}_z & B_z^{(20)} & B_z^{(10)} & B_z^{(5)} & B_z^{(2)} & B_z^{(1)} \\ \hline
A1v &  109 &   55 &  0.27 &  0.00 &  0.05 &  0.13 &  0.31 &  0.52 &  1.48 &  2.33 \\ % m288a5
A2v &  105 &   52 &  0.35 &  0.00 &  0.10 &  0.19 &  0.51 &  0.81 &  1.83 &  2.64 \\ % m288a4
A3v &   94 &   47 &  0.39 &  0.00 &  0.25 &  0.36 &  0.78 &  1.11 &  2.11 &  2.76 \\ % m288a1
A4v &   83 &   42 &  0.36 &  0.00 &  0.49 &  0.61 &  1.03 &  1.41 &  2.37 &  3.02 \\ % m288a2
A5v &   74 &   37 &  0.33 &  0.00 &  0.74 &  0.83 &  1.35 &  1.75 &  2.63 &  3.13 \\ % m288a6
A6v &   68 &   34 &  0.30 &  0.00 &  0.99 &  1.06 &  1.59 &  2.05 &  2.91 &  3.37 \\ % m288a3
\hline
A1h &  114 &   46 &  0.12 &  0.05 &  0.00 &  0.01 &  0.05 &  0.10 &  0.40 &  0.97 \\ % m288b4
A2h &  110 &   44 &  0.22 &  0.12 &  0.00 &  0.03 &  0.09 &  0.16 &  0.59 &  1.40 \\ % m288b2
A3h &  103 &   41 &  0.31 &  0.25 &  0.00 &  0.05 &  0.18 &  0.32 &  1.01 &  1.98 \\ % m288b1
A4h &   90 &   30 &  0.36 &  0.49 &  0.00 &  0.12 &  0.33 &  0.63 &  1.48 &  2.43 \\ % m288b3
A5h &   76 &   25 &  0.26 &  0.99 &  0.00 &  0.19 &  0.59 &  0.82 &  1.81 &  2.57 \\ % m288b5
\hline
B1v & \mbox{LES}&   51 &  0.50 &  0.00 &  0.09 &  0.70 &  1.11 &  2.01 &  3.46 &  3.91 \\ % ms512c1_rss
B2v & \mbox{LES}&   50 &  0.58 &  0.00 &  0.17 &  0.88 &  1.31 &  2.25 &  3.57 &  4.06 \\ % ms512c2_rss
B3v & \mbox{LES}&   44 &  0.65 &  0.00 &  0.45 &  1.25 &  1.60 &  2.50 &  3.74 &  4.17 \\ % ms512c3_rss
B4v & \mbox{LES}&   37 &  0.53 &  0.00 &  0.86 &  1.37 &  1.94 &  2.73 &  3.89 &  4.29 \\ % ms512c4_rss
\hline
C1v & \mbox{LES}&   76 &  0.82 &  0.00 &  0.23 &  1.83 &  2.83 &  3.68 &  4.18 &  4.23 \\ % ms1024a1_rss
C2v & \mbox{LES}&   69 &  0.85 &  0.00 &  0.46 &  1.93 &  2.97 &  3.80 &  4.22 &  4.26 \\ % ms1024a2_rss
C3v & \mbox{LES}&   59 &  0.68 &  0.00 &  0.92 &  2.11 &  3.18 &  3.93 &  4.30 &  4.34 \\ % ms1024a3_rss
\hline
C1h & \mbox{LES}&   79 &  0.80 &  0.23 &  0.00 &  0.05 &  0.14 &  0.36 &  1.10 &  2.09 \\ % ms1024b1_pc
C2h & \mbox{LES}&   80 &  0.60 &  0.23 &  0.00 &  0.20 &  0.47 &  1.15 &  2.90 &  3.77 \\ % ms1024b1_rss
C3h & \mbox{LES}&   52 &  0.63 &  0.46 &  0.00 &  0.64 &  1.40 &  2.64 &  3.78 &  4.03 \\ % ms1024b2_rss
C4h & \mbox{LES}&   34 &  0.42 &  0.92 &  0.00 &  1.08 &  2.00 &  3.16 &  3.88 &  4.01 \\ % ms1024b3_rss
            \hline
          \end{array}
          $$
\tablefoot{LES in the column for $\Rey$ indicates runs where
  Smagorinsky viscosity is used. We apply vertical field conditions
  for the magnetic field in all runs except C1h where the top boundary
  is perfectly conducting. The data in the last seven
  columns are given in units of kG. The last five columns refer to
  temporally averaged maxima of low-pass filtered vertical magnetic
  field $B_z$ at a depth of roughly 1~Mm, and where the superscripts
  1, 2, 5, 10, and 20 refer to the filtering scale in Mm.}
\end{table}

\subsection{Application to solar parameters}
\label{sec:solpar}

To compare with the Sun, it is convenient to
transform the results into physical units. This can be done in several
ways, which can place the computational domain at different depths in
the solar convection zone.
As the sunspot are manifestations of the solar magnetic field at the
surface, it is logical to place the computational domain near the
surface.
We assume that the pressure scale height, gas density, and the temperature
at the surface are the same as in the Sun,
i.e.\ $\Hp^{(\odot)}\approx1.5\cdot 10^5$~m,
$\rho_\odot=2.5\cdot10^{-4}$~kg~m$^{-3}$, and $T_\odot=5800$~K,
respectively, defining the units
of length, density, and temperature. Furthermore, we take the
acceleration due to
gravity to have the solar surface value $g_\odot=274$~m~s$^{-2}$, and use
the permeability of vacuum, $\mu_0 = 4\pi\cdot10^{-7}$~N~A$^{-2}$ to
derive the unit of magnetic field.
With these choices we obtain units
\begin{eqnarray}
\left[x\right] &=& \Hp^{(\rm cool)} = \Hp^{(\odot)},\\
\left[t\right] &=& (\Hp^{(\rm cool)}/g)^{1/2} = (\Hp^{(\odot)}/g_\odot)^{1/2},\\
\left[\rho\right] &=& \rho_{\rm top} = \rho_\odot,\\
\left[T\right] &=& T_{\rm cool} = T_\odot,\\
\left[B\right] &=& \left(\mu_0 \rho_{\rm top} g \Hp^{(\rm cool)}\right)^{1/2}
= \left(\mu_0 \rho_\odot g_\odot \Hp^{(\odot)}\right)^{1/2},
\end{eqnarray}
where $\rho_{\rm top}=\rho(z=0)$ is the surface density,
while $\Hp^{(\rm cool)}$ and $T_{\rm cool}$ are the
pressure scale height and temperature in the cooling layer, respectively.

The profiles of horizontally averaged root-mean-square velocity and
the equipartition magnetic field strength $B_{\rm eq}=\langle\mu_0
\rho {\bm u}^2\rangle^{1/2}$ from the hydrodynamic progenitor runs for
each of our density stratifications are shown in
Fig.~\ref{fig:purms}.
The depths of the domains are now 8.5~Mm in Set~A, 12.5~Mm in Set~B,
and 24~Mm in Set~C with horizontal sizes of 34, 50, and 96~Mm,
respectively. The box in our Set~C is comparable to the domain
size used by \cite{SN12}.
We find that the velocities near the surface are of the order of
$2$--$3$~km~s$^{-1}$, which is similar to the convective velocities
observed in the Sun and also obtained from mixing length theory
\citep[e.g.][]{Stix02}. The
lower overall velocity in Run~C00 is due to a lower input energy flux
in comparison to the other runs due to a lower value of $K$ which was
adopted in order to limit the Mach number near the surface. We note
that, using the mixing length model of \cite{Stix02}, we obtain a value
of $\mathcal{F}\approx2.7\cdot10^{-7}$ in the Sun at a depth of
roughly 24~Mm. The
equipartition magnetic field strength is of the order of 3~kG in
Sets~B and C. The lower value in Set~A is due to the overall lower
density in the interior for the runs in that set.

Using these values, the imposed magnetic field strength in Set~C, where
the most clear indications of flux concentrations are visible, is in
the range $230$--$920$~G, see Table~\ref{tab:runs}. The maximum
strength of the concentrations shown, e.g.\ in Figs.~\ref{fig:bb3_xy2}, and
\ref{fig:pcross_section_ms1024a1}, is in the range $3$--$5$~kG and the
size of the largest field concentrations in our simulations are of the
order of 20~Mm. Both of these are in the range observed for sunspots.

\section{Results} \label{sect:results}

We perform three sets of simulations in which we increase the size of
the domain systematically while keeping the box aspect ratio fixed,
see Table~\ref{tab:sets}.
We study the cases of horizontal and vertical imposed fields and
analyse the detected flux concentrations separately for the two
cases. We also measure the effective magnetic pressure from all runs
and study whether NEMPI can be the explanation for the observed
features.

\begin{figure*}[t]
\centering
\includegraphics[width=0.33\textwidth]{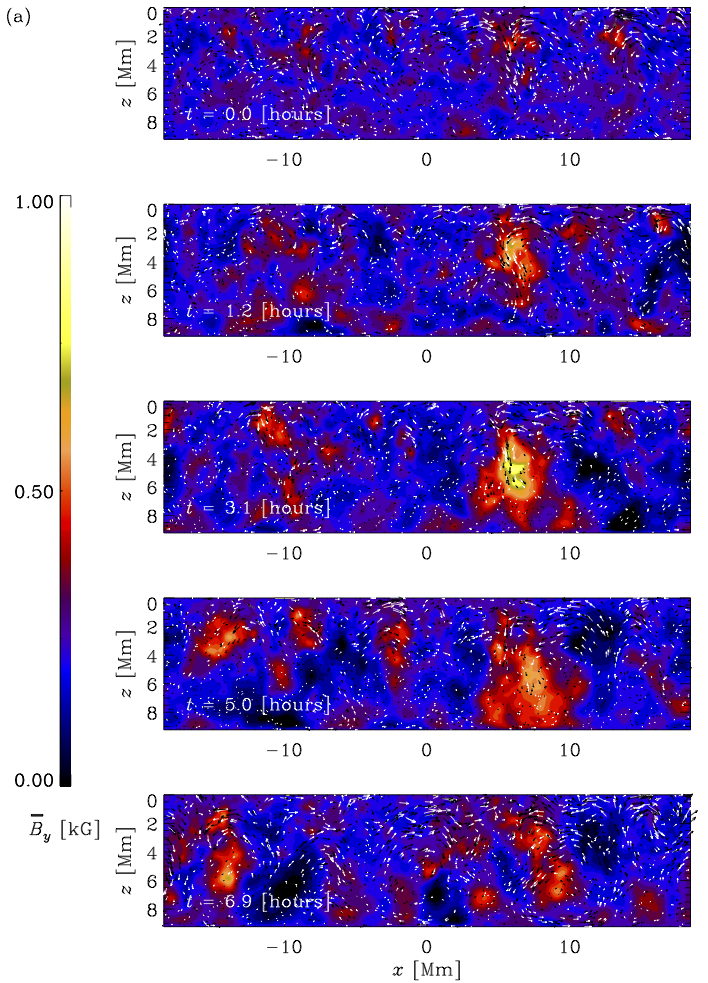}
\includegraphics[width=0.33\textwidth]{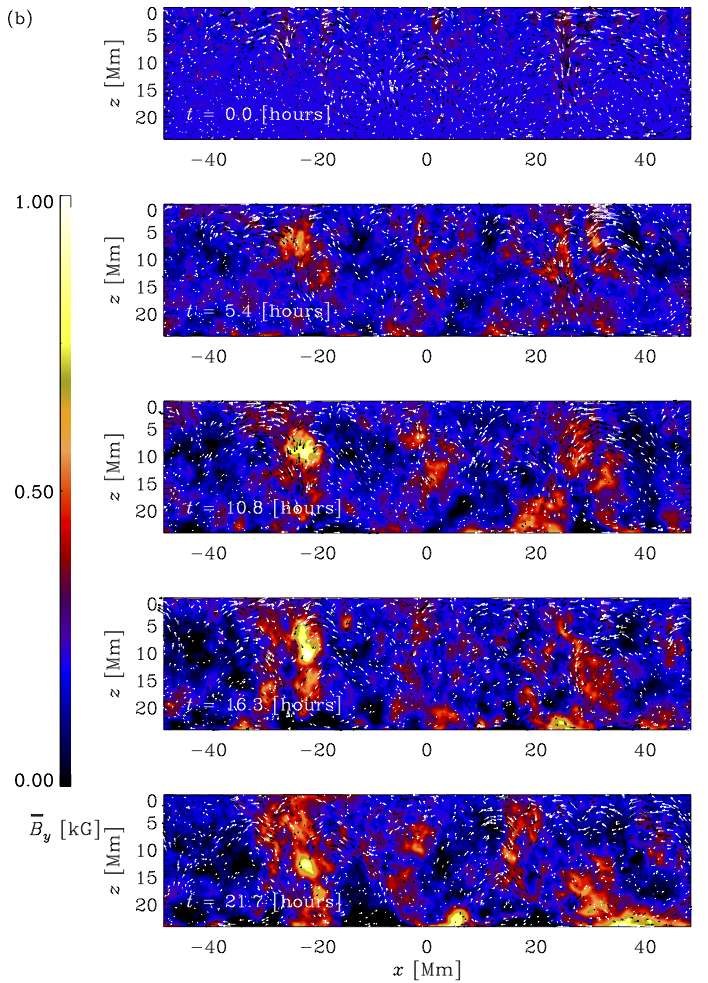}
\includegraphics[width=0.33\textwidth]{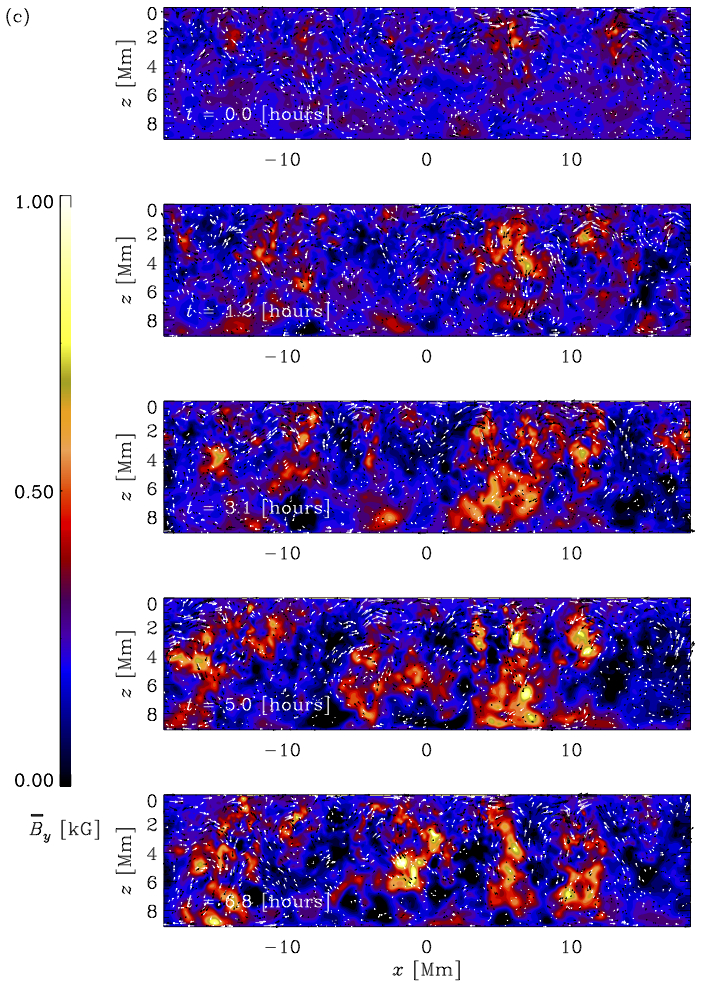}
\caption{(a) Mean magnetic field component
  $\mean{B}_y=\mean{B}_y(x,z)-\mean{B}_y(z)+B_0$
  in units of kG from Run~A3h from five different times indicated in the
  legends.
  (b) Same as \Fig{fig:pby_xz_m288b1}, but for Run~C1h.
  (c) The same as Fig.~\ref{fig:pby_xz_m288b1}(a) but from an otherwise
  similar run, except
  where the Lorentz force and Ohmic heating are omitted.
  The white and black arrows indicate the $y$-averaged flows in the
  $(x,z)$ plane.
}
\label{fig:pby_xz_m288b1}
\end{figure*}

\begin{figure*}[t]
\centering
\includegraphics[width=0.33\textwidth]{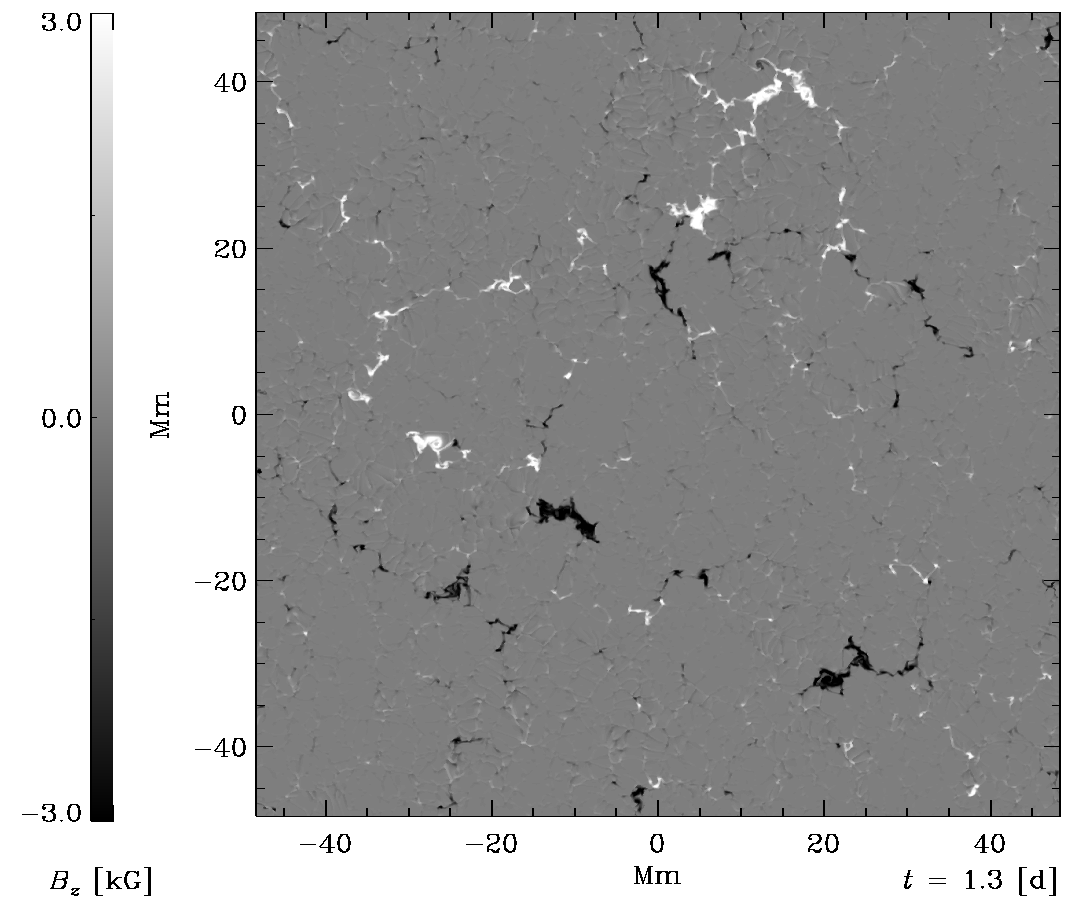}
\includegraphics[width=0.33\textwidth]{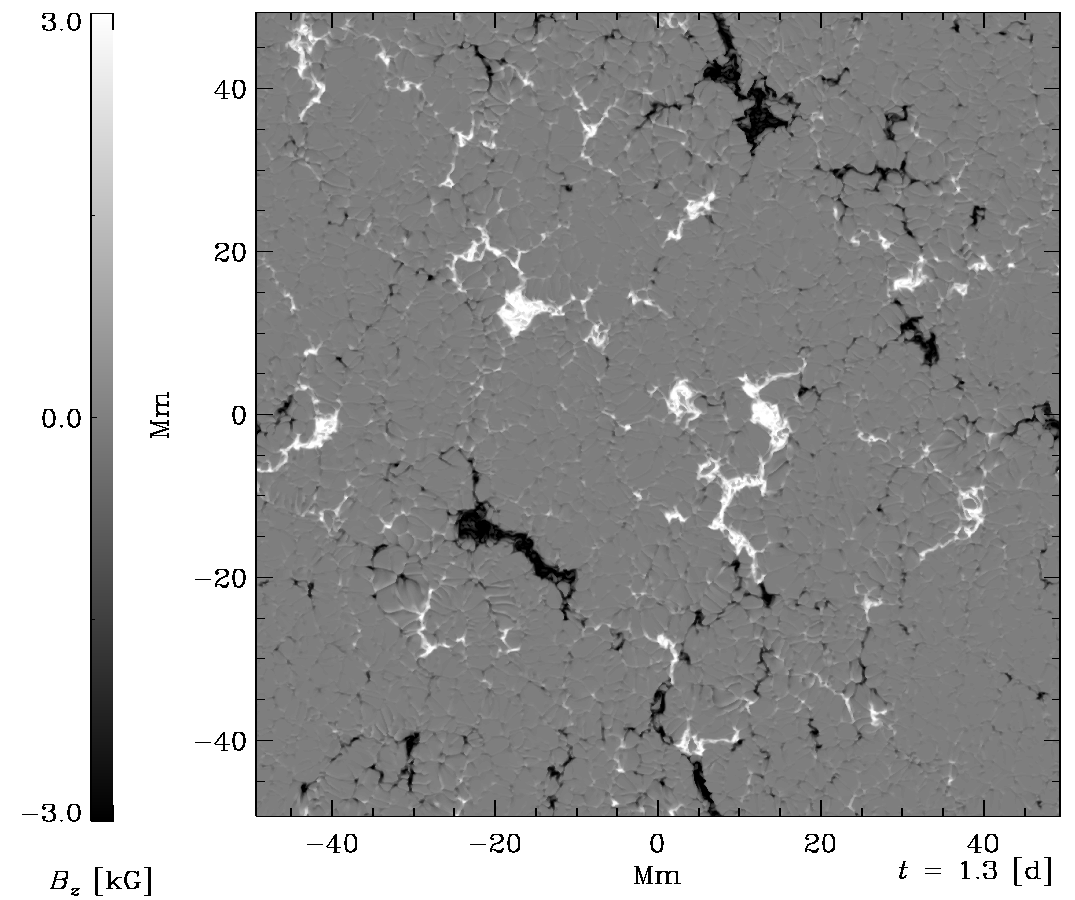}
\includegraphics[width=0.33\textwidth]{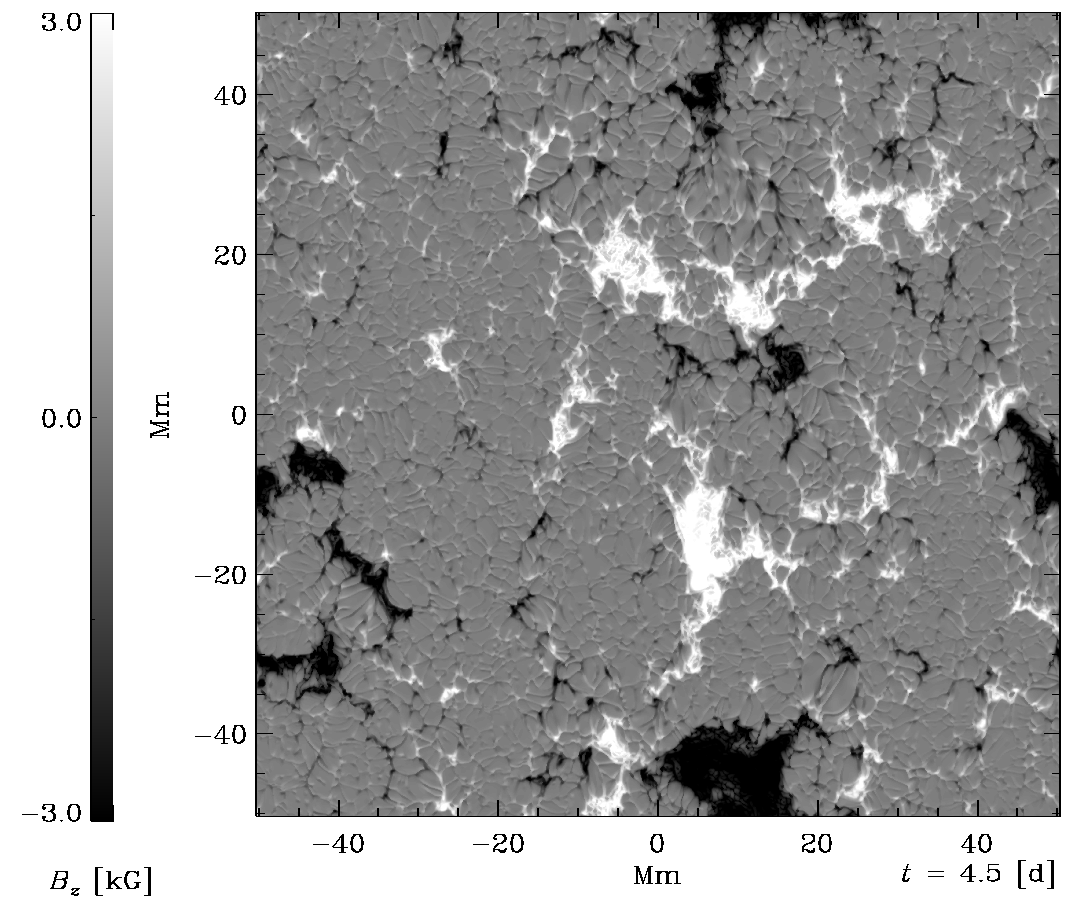}
\caption{Vertical magnetic field $B_z$ near the surface at a depth of
  0.6~Mm from representative snapshots of Runs~C2h (left panel), C3h
  (middle), and C4h (right). The magnetic field scale is clipped at
  $\pm3$~kG in each panel. The maximum field strengths obtained are of
  the order of 5~kG. Animation associated with Run~C3h can be found
  here: \url{http://research.ics.aalto.fi/cmdaa/group-Movies.shtml}.}
\label{fig:bb3_xy2}
\end{figure*}

\subsection{Imposed horizontal field}

Early studies of negative effective magnetic pressure
and NEMPI in turbulent convection have been performed with an imposed
{\em horizontal} field \citep{KBKMR12,KBKMR13}.
This choice is motivated by the anticipated presence of a diffuse, azimuthally
dominated large-scale field in the bulk of the solar convection
zone, the origin of which is e.g.\ an $\alpha\Omega$-type dynamo
\citep{WKKB14}.
When NEMPI is excited, magnetic field concentrations were best detected
in averages taken along the direction of the imposed field
\citep{BKKMR11,KeBKMR12,KeBKMR13}
if the scale separation between forcing scale and the size of the box
is smaller than 30.
We show two such cases for Runs~A3h and C1h with the lowest and highest
stratifications in Figs.~\ref{fig:pby_xz_m288b1}(a) and
\ref{fig:pby_xz_m288b1}(b), respectively. We find flux
concentrations with maximum field strength or the order of 1~kG, which
is roughly four times the imposed field strength. This is similar to what
was obtained in the aforementioned studies employing forced turbulence
clearly showing NEMPI.

In the present case, the flux concentrations
are associated with large-scale downflows (see the black/white arrows
in Fig.~\ref{fig:pby_xz_m288b1}).
The concentrations become visible near the surface in regions of
converging flows. In the 8.5~Mm domain the structures descend to a
depth of roughly 6~Mm in five hours, see
Fig.~\ref{fig:pby_xz_m288b1}(a). The timescale in Run~C1h appears
similar, see the second panel from the top of
Fig.~\ref{fig:pby_xz_m288b1}(b), and the concentration reaches the
bottom of the domain in roughly 25 hours, corresponding to roughly ten
large-scale convective turnover times.
This is similar to the so-called ``potato-sack'' effect where horizontal
magnetic structures become
heavier than their surroundings, often observed as a consequence of
the negative effective magnetic pressure.
This effect was found in both DNS and mean-field simulations (MFS) of forced
turbulence \citep{BKKMR11,KeBKMR13}, where the downflows of the magnetic concentrations
can be directly associated with
the negative effective magnetic pressure.
In turbulent convection, the potato-sack effect was previously found only
in MFS \citep{KBKMR12}. In the present study we detect a similar
effect for the first time in DNS and LES of convection;
see Figs.~\ref{fig:pby_xz_m288b1}(a) and~\ref{fig:pby_xz_m288b1}(b).
On the other hand, in convection, downflows occur naturally without the presence of
the negative effective magnetic pressure,
so it is not a priori clear whether these downflows are affected or
even driven by the magnetic field, as was found in isothermal forced
turbulence, where no thermal buoyancy is possible.

\begin{figure*}[t]
\centering
\includegraphics[width=\textwidth]{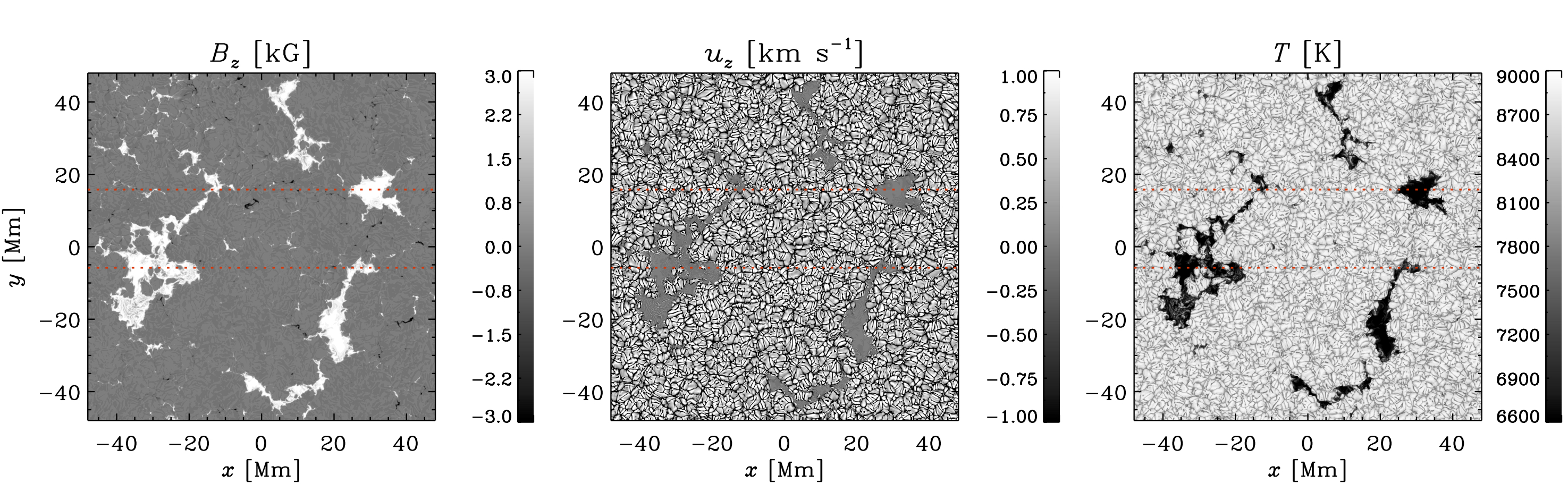}
\includegraphics[width=\textwidth]{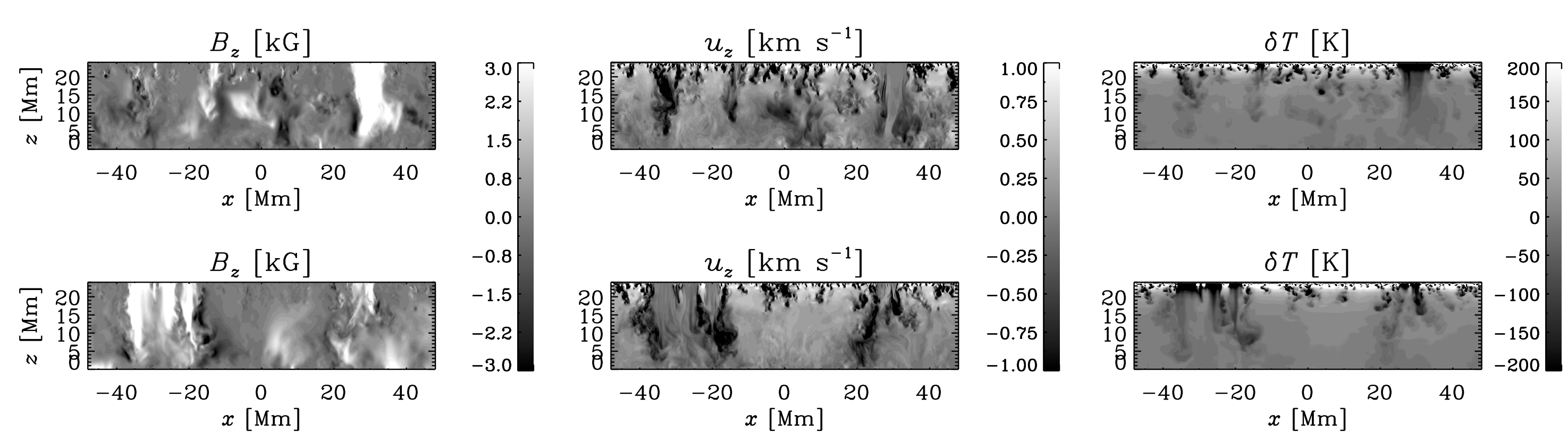}
\includegraphics[width=\textwidth]{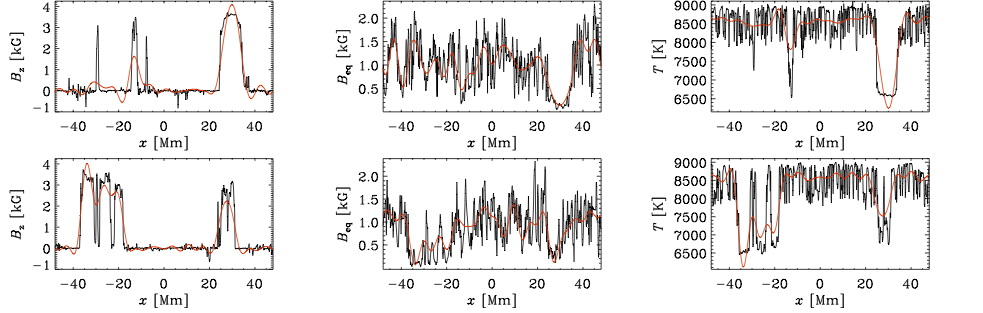}
\caption{Top row: vertical magnetic field $B_z$, vertical velocity
  $u_z$, and temperature $T$, respectively, at $z=0.6$~Mm for Run~C1v.
  The second and third rows show vertical
  cuts from cuts through $y=15.9$~Mm and $y=-5.8$~Mm. In the rightmost
  panels we show the $\delta T = T - \mean{T}(z)$ and oversaturate the
  scale so that structures in the deeper layers become visible.  The
  line plots on the last two rows show the vertical magnetic field and
  equipartition field strength, and temperature at $z=0.6$~Mm from the
  same $y$-positions. The black lines in the middle and bottom rows
  show The red lines indicate
  low-pass filtered data where the filtering scale is $d_{\rm
    sm}=6.0$~Mm. The positions of the cuts are indicated as red dotted
  lines in the uppermost row. Animation associated with this run can
  be found here: \url{http://research.ics.aalto.fi/cmdaa/group-Movies.shtml}.}
\label{fig:pcross_section_ms1024a1}
\end{figure*}

As a control, we run one of the models (Run~A3h) from the same
initially hydrodynamic snapshot and neglect the Lorentz-force and Ohmic
heating. In this simulation the induction equation does not affect the
flow and the magnetic field is a passive vector. We show
in Fig.~\ref{fig:pby_xz_m288b1}(c)
the passive vector evolution, corresponding to the magnetic field
evolution in Fig.~\ref{fig:pby_xz_m288b1}(a).
We find that a flux concentration
forms near $x\approx14.5$~Mm, similarly as in the MHD run. This is
explained by a downflow that existed already in the hydrodynamic parent
run.
However, in the passive vector case, the concentration is somewhat
weaker, less coherent, and the time
scale after which the structure reaches the bottom of the convection zone
is shorter.
The latter is likely a consequence of missing magnetic buoyancy in
the passive vector model.
Thus it appears that the downflows, although
characteristic of the formation of magnetic concentrations,
are already present in the hydrodynamic case and play a crucial role
in concentrating the flux. We discuss the role of the negative
effective magnetic pressure in Sect.~\ref{sec:peff}.

In the earlier simulations of magnetic flux concentrations in stratified convection with an
imposed horizontal field \citep{KBKMR12,KBKMR13} a perfect conductor
boundary condition did not allow the formation of spot-like structures
near the surface.
However, in highly stratified simulations when potential or vertical
field conditions are applied, the studies of \cite{SN12} and
\cite{WLBKR13} found the possibility of bipolar-region
formation.
Motivated by these results we apply a
vertical field condition in most of the current models.
The surface appearance of the magnetic fields of Runs~C1h--C3h is shown in
Fig.~\ref{fig:bb3_xy2}. For the weakest imposed field (Run~C1h,
$|\BBB_0|\approx230$~G $\approx0.07 \Beq$)
we find rather small concentrations of
either sign, but no clear bipolar regions. As the imposed field
strength is increased, the size of the concentrations grows. In the
case with the strongest imposed field (Run~C3h, where
$|\BBB_0|\approx920$~G $\approx0.38 \Beq$),
the maximum horizontal size of the surface
structures is roughly $20$~Mm, and it is possible to identify bipolar
spot pairs. To quantify this we study low-pass filtered data of $B_z$
from slices taken near the surface. We apply five filtering scales
between 1 and 20~Mm, see Table~\ref{tab:runs}. We find that the
maximum field strength in the case where the smallest retained scale
is 20~Mm increases from 0.06 in Run~A3h to 0.21 in Run~C2h. The
maximum field strength in the two largest scales ($B_z^{(10)}$ and
$B_z^{(20)}$) increases roughly proportionally to the imposed field
strength in Sets~A and C, see the 6th and 7th columns of
Table~\ref{tab:runs}, indicating the presence of large-scale magnetic
structures. The increase in the cases of smaller filtering scales is
less dramatic, especially in Set~C with the larger domain size.

\begin{figure*}[t]
\centering
\includegraphics[width=\textwidth]{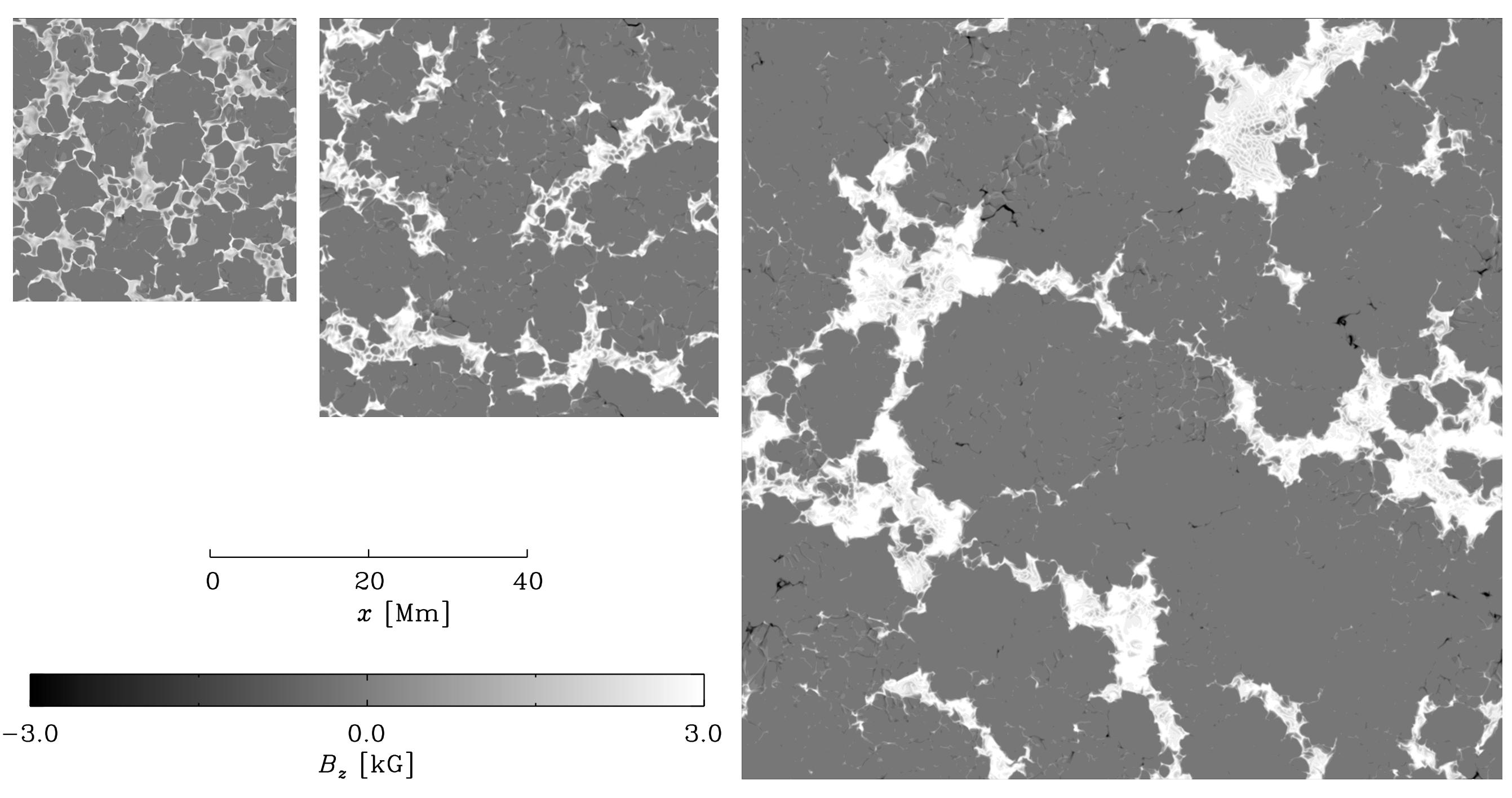}
\caption{Horizontal slices of $B_z$ near the surface from Runs~A4v,
  B2v, and C2v with different box sizes. The physical scale is shown
  in the legend.}
\label{fig:pslices_hori}
\end{figure*}

\subsection{Imposed vertical field}

Pronounced effects of the negative effective magnetic pressure
have been found in the case of an imposed vertical field in studies
where turbulence is forced \cite[e.g.][]{BKR13,BGJKR14,LBKR14}.
This is because, unlike a horizontal field, a vertical one is not
advected by the resulting downflow,
i.e., there is no potato-sack effect.
However, as the downflow removes gas from the upper layers, the pressure
decreases, which results in a return flow that draws with it more
vertical field.
This can lead to field amplification to a strength that exceeds the
equipartition field strength in the top layers; see \cite{BKR13} for
numerical simulations in isothermal stratified turbulence.
In the aforementioned studies the field concentrations often form a
spot-like structure because the ratio between the domain size and
forcing scale is sufficiently large \cite[e.g.][]{BKR13,BGJKR14,LBKR14}.

\begin{figure}[t]
\centering
\includegraphics[width=\columnwidth]{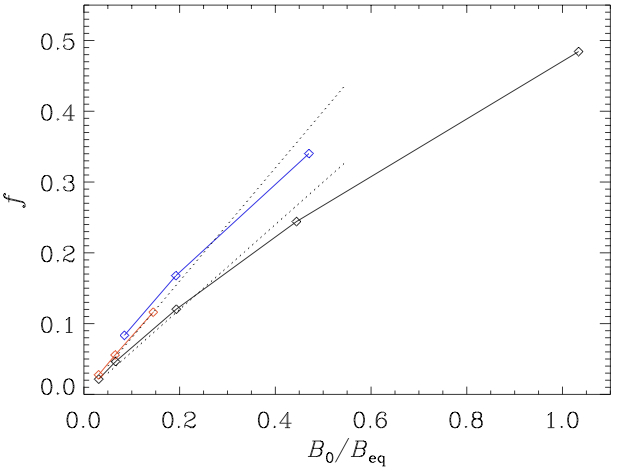}
\caption{
The magnetic field redistribution factor (the relative areas in which
vertical fields exceeding the equipartition value, $B_z > \Beq$)
in runs with vertical fields
from Sets~A (black), B (red), and C (blue). The dotted lines are
proportional to $\Beq$.}
\label{fig:pfillingf}
\end{figure}

\begin{figure*}[t]
\centering
\includegraphics[width=\textwidth]{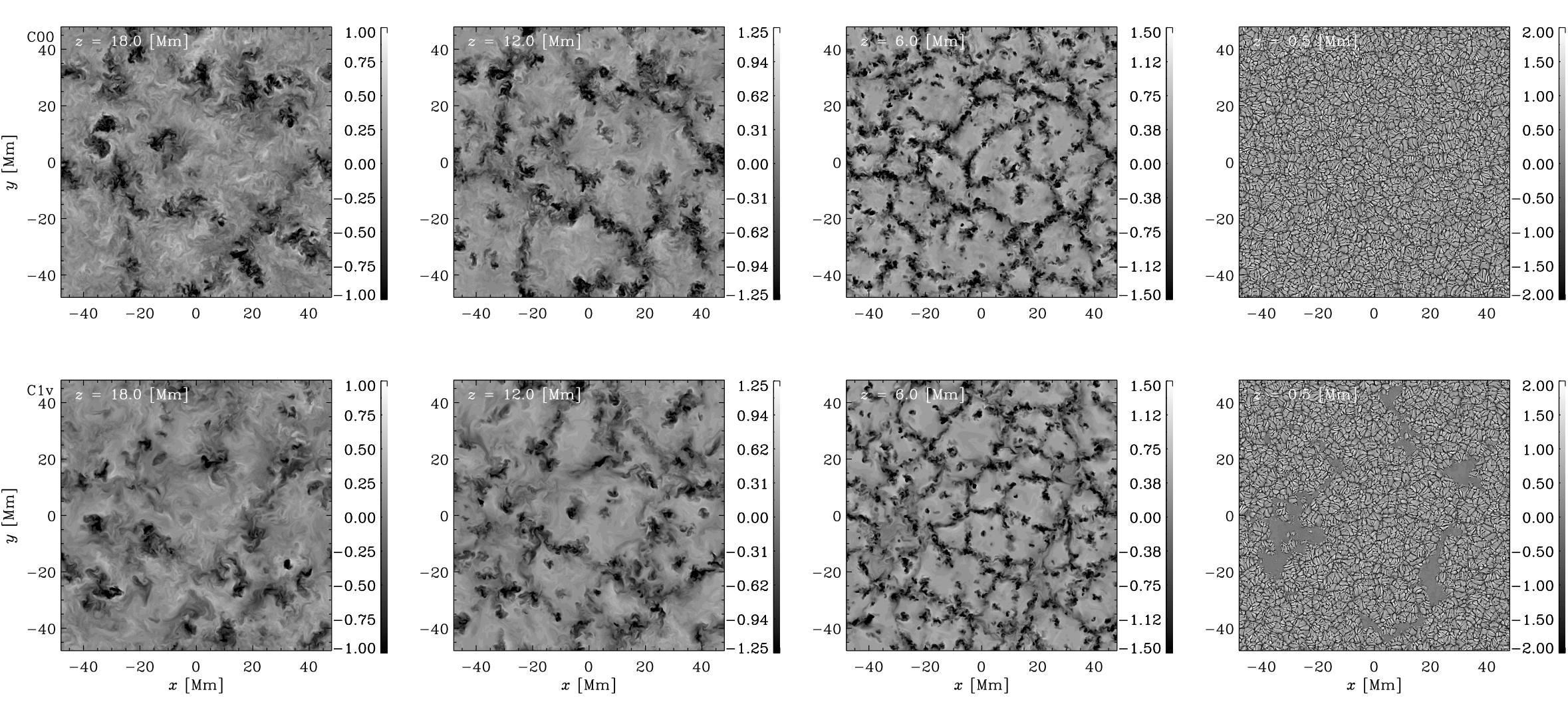}
\caption{Vertical velocity $u_z$ from four depths as indicated in the
  legends from a hydrodynamical run C00 (top row) and a run with
  imposed vertical field C1v (bottom row). The velocity is given in
  units of km~s$^{-1}$.}
\label{fig:puu_slices_ms1024a1}
\end{figure*}

In the top row of \Fig{fig:pcross_section_ms1024a1} we show
visualizations of the
vertical magnetic field $B_z$, velocity $u_z$, and temperature $T$
from a depth of $0.6$~Mm for Run~C1v with an imposed vertical field of
$230$~G.
Note that there are now three large patches, the largest exceeding
20~Mm in diameter, where positive $B_z$ of the order of 3~kG is
found. Line plots through two of the patches (two bottom panels of
\Fig{fig:pcross_section_ms1024a1}) show that the magnetic field
exceeds the local equipartition field strength by a factor
of more than ten due to the fact that convection is nearly completely
suppressed in regions of strong magnetic fields.
The temperature within the magnetic structures at the depth of
$0.6$~Mm is reduced by roughly 2000~K which is within the observed
range for sunspots. We also find that the structures penetrate almost
the entire depth of the layer, see the second and third rows of
\Fig{fig:pcross_section_ms1024a1}. The temperature is affected mostly
near the surface, whereas in the deeper layers the difference to the
ambient atmosphere is 1--2 orders of magnitude smaller than near
the surface.
The structures are qualitatively similar to those seen in forced
turbulence simulations with poor scale separation, where they are
caused by NEMPI, see Fig.~17 of \cite{BGJKR14}.
This result is also reminiscent of early work of \cite{TWBP98}, who found similar
behavior in large aspect ratio convection simulations, although at much
smaller Rayleigh numbers and weaker density stratification.

Representative results of the vertical field near the surface from the
three domain sizes with comparable imposed fields of the order of
0.5~kG are shown in Fig.~\ref{fig:pslices_hori}. We find that the size
of the structures increases from roughly 5~Mm to 20~Mm as the domain
size is increased from 34~Mm to 96~Mm. Also the field topology
changes from a web-like network of strong fields in Run~A4v with the
smallest domain size to one with more isolated structures in Run~C2v for
the largest physical size. A possible explanation is that the
equipartition field is smaller in Run~A4v (see Fig.~\ref{fig:purms})
than in the other two runs and that the magnetic field has a greater
effect on the flow. A similar transition from isolated magnetic
structures for relatively weak fields to a network-like structure for
intermediate field strengths has previously been reported by
\cite[e.g.][]{TWBP98,TP13}. We have not explored such strong fields as
\cite{TP13}, which would induce small-scale convection throughout the
domain, as seen in the flux concentrations in the rightmost panel of
Fig.~\ref{fig:pslices_hori}.

The magnetic field redistribution factor (the relative areas in which
vertical field exceeds the equipartition value)
is roughly proportional to the imposed field strength, see
Fig.~\ref{fig:pfillingf}.
This result follows from the conservation law for the total
magnetic flux, $B_0 \hat S= \Beq \hat S_1 + (\hat S-\hat S_1) B_{\rm res}$,
where $\hat S$ is the total area and $\hat S_1$ is the area of the strong field (about
the equipartition field), $B_{\rm res} \ll \Beq$ is the final weak magnetic field
(much smaller than the equipartition field).
This yields $f=\hat S_1/\hat S \propto B_0/\Beq$.

As in the case of the imposed horizontal field, we find here for the
vertical field that the large-scale contribution indicative of
magnetic flux concentrations, increases as the imposed field strength
is increased (see Table~\ref{tab:runs}). The growth of the maximum,
however, is significantly less steep in the vertical field case
especially in Sets~B and C. In Set~C, $B_z^{(20)}$ increases only by 20
per cent when the imposed field increases fourfold.

Given that the negative effective magnetic pressure
is capable of producing downflows in ways similar
to thermal convection, one wonders whether there are any noticeable
differences between downflows produced with and without magnetic fields.
\Fig{fig:puu_slices_ms1024a1} shows a comparison between the two
(Runs~C00 and C1v).
For horizontal fields we have seen above that in both cases there are
downflows, but it is not clear whether they are significantly affected
by the presence of flux concentrations.
Here, the most pronounced difference occurs immediately in the top layer,
where we see large-scale patches with almost vanishing velocity in the
areas where strong magnetic fields are present.
Some extended patches are also still seen at a depth of $z=6$~Mm, but
they are now subdominant compared with the narrower downdrafts.
In deeper layers, however, below $z=12$~Mm, the flow structure
is rather similar in both cases, except that in the case with
magnetic field the flow patterns are somewhat smoother.
A similar effect of dynamo-generated magnetic fields on the
small-scale flow structure has been noted by \cite{HRY15b}.

\begin{figure}[t]
\centering
\includegraphics[width=\columnwidth]{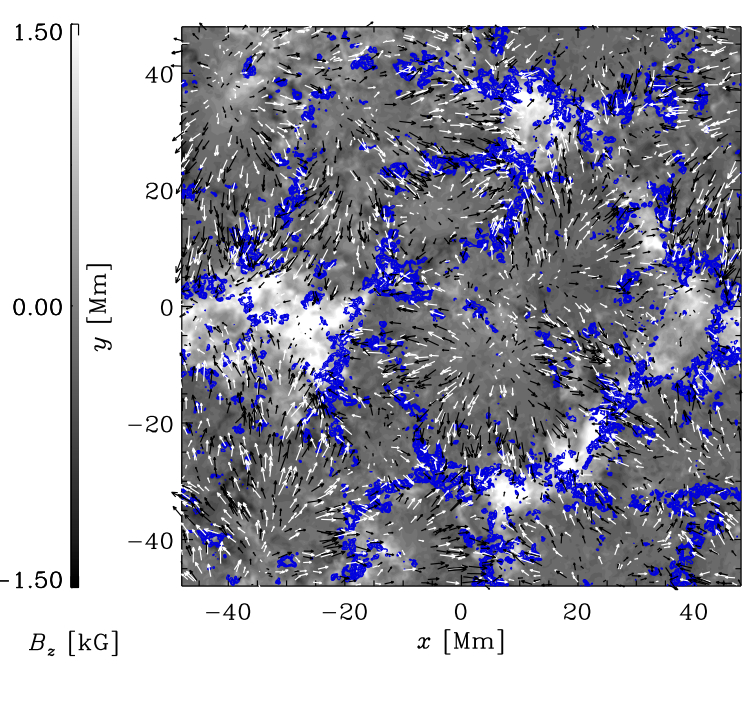}
\caption{Temporally averaged vertical magnetic field (black and white
  contours), horizontal flows (black and white arrows), and downflows
  exceeding 250~m~s~$^{-1}$ (blue contours) at a depth of 6~Mm in Run~C1v.}
\label{fig:pbuav}
\end{figure}

We find that the magnetic concentrations tend to appear in regions
where large-scale convective downflows occur, see Fig.~\ref{fig:pbuav}
in which the temporally averaged vertical magnetic field is shown along
with the similarly averaged flows from Run~C1v. Extracting the
large-scale fields was obtained by temporally averaging over ten
snapshots, each separated by 4.5 hours, of the simulation data. The
horizontal scale of the large-scale cells is roughly 40--50~Mm, and
they span the entire vertical extent of the domain. Flows at these
scales correspond to supergranulation in the Sun. The fact that the
flux concentrations are situated at the vertices of the large-scale
convection pattern suggests that their origin is the flux expulsion
mechanism proposed by \cite{W66}.

\begin{figure}[t]
\centering
\includegraphics[width=\columnwidth]{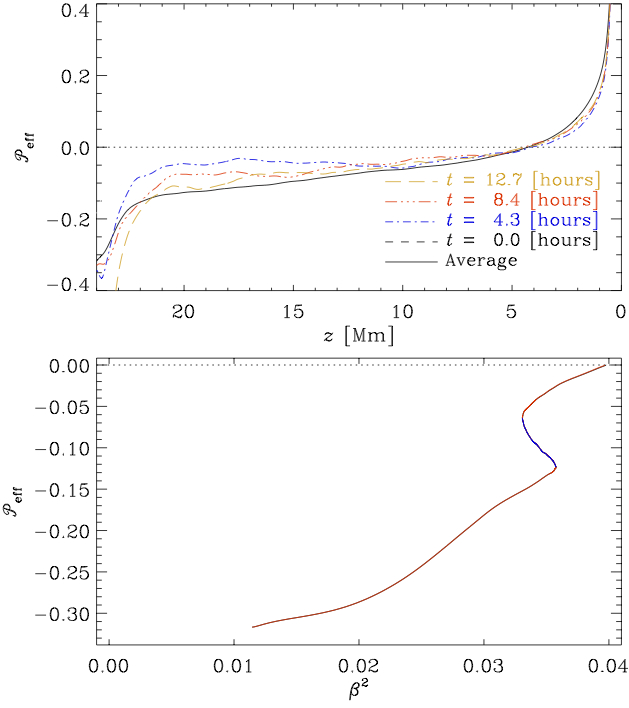}
\caption{Top panel: Effective magnetic pressure $\Peff$ as a function of height
  for Run~C2v. The solid black line shows the time averaged data,
  whereas the other curves show instantaneous data from times
  indicated in the legend. Bottom panel: $\Peff$ as a function of
  $\beta^2$ in regions where $\Peff<0$ for the temporally averaged
  data from the top panel. Red (blue) part of the curve indicates
  $d\Peff/d\beta^2>0$ ($d\Peff/d\beta^2<0$).}
\label{fig:pPeff_prof_ms1024a2_rss}
\end{figure}

\begin{figure}[t]
\centering
\includegraphics[width=\columnwidth]{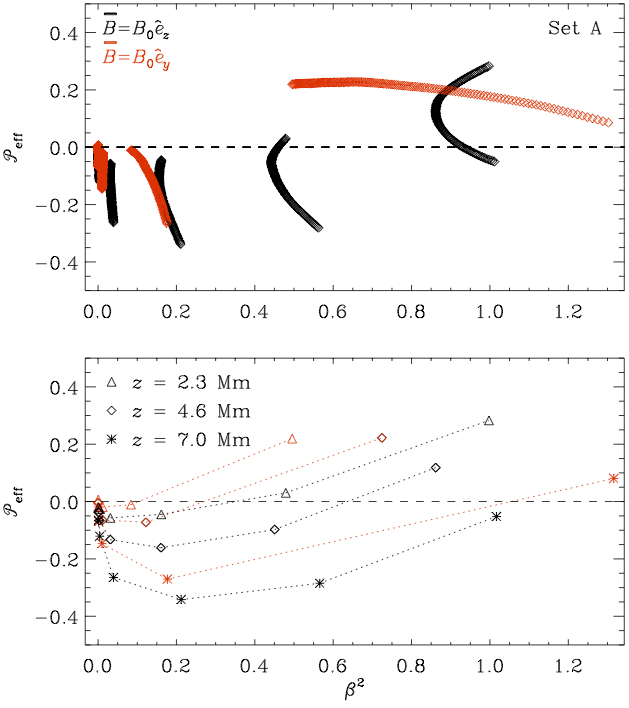}
\caption{Top panel: mean effective magnetic pressure as a function of
  $\beta$ for the runs in Set~A with vertical (black) and horizontal
  (red) imposed fields from the range $2.3~\mbox{Mm} \leq z \leq
  7.0~\mbox{Mm}$. Lower panel: $\Peff$ at heights $z=2.3$~Mm
  (triangles), $4.6$~Mm (diamonds), and $7.0$~Mm (stars).}
\label{fig:pPeff_all}
\end{figure}

\begin{figure}[t]
\centering
\includegraphics[width=\columnwidth]{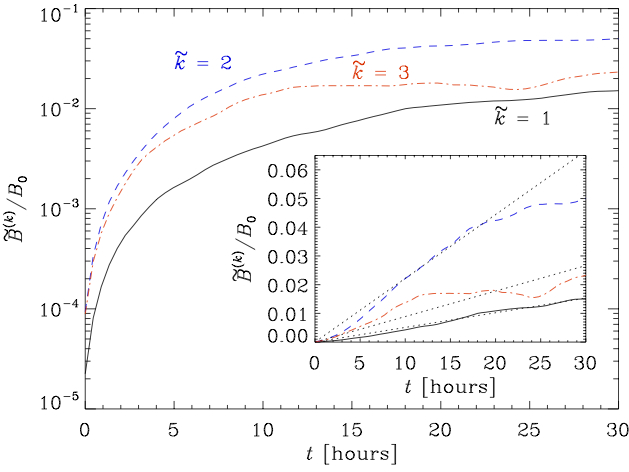}
\includegraphics[width=\columnwidth]{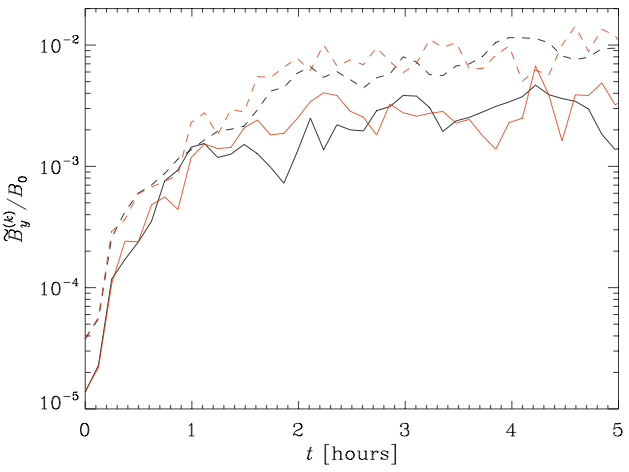}
\caption{Top panel: Normalised Fourier amplitudes
  $\tilde{B}^{(k)}/B_0$ for the
  wavenumbers $\tilde{k}=k/k_1=1\ldots3$, where $k_1=2\pi/L_x$, as
  functions of time from depth of $0.6$~Mm in
  Run~C1v. The inset shows the same in linear scale. The dotted lines
  in the inset indicate growth linearly proportional to time. Bottom
  panel: Fourier amplitudes of $\tilde{k}=1$ (solid lines) and $2$
  (dashed lines) for runs A3h (black lines) and a corresponding runs
  without backreaction from the magnetic field to the flow (red
  lines).}
\label{fig:pBB3_growth_ms1024a1_rss}
\end{figure}

\subsection{Effective magnetic pressure}
\label{sec:peff}

In our study we measure the effective magnetic pressure in order to
clarify the role of NEMPI in the formation of inhomogeneous magnetic structures
in turbulent convection.
Below we define the effective magnetic pressure and describe
the method of its measurement.
The total turbulent stress, including the contributions of Reynolds
and Maxwell stresses is given by
\begin{eqnarray}
\Pi_{ij}^{\rm (f)}=\mean{\rho} \mean{u_iu_j} + \delta_{ij}
\mean{\bm b}^2/2\mu_0 - \mean{b_ib_j}/\mu_0,
\end{eqnarray}
where $\delta_{ij}$ is the Kronecker tensor and the superscript `(f)'
refers to contributions from the fluctuations.
The turbulent stress is
split into two contributions that are either independent
($\Pi_{ij}^{\rm (f,0)}$) of or dependent ($\Pi_{ij}^{{\rm
    (f,}\meanv{B}{\rm )}}$) on the mean field. Their difference
$\Delta\Pi_{ij}^{\rm (f)} = \Pi_{ij}^{{\rm (f,}\meanv{B}{\rm
    )}}-\Pi_{ij}^{\rm (f,0)}$ is due to
the mean magnetic field and can be parametrised in the form
\begin{eqnarray}
\Delta\Pi_{ij}^{\rm (f)}=\mu_0^{-1} \left(q_s\mean{B}_i\mean{B}_j
- {1 \over 2} q_p \delta_{ij} \meanv{B}^2 - q_g \hat{g}_i\hat{g}_j \meanv{B}^2\right),
\label{equ:dPij}
\end{eqnarray}
where $\hat{g}_i$ is the unit vector along the direction of gravity.
Furthermore, $q_s$ represents the contribution of turbulence to the mean
magnetic tension and $q_p$ is the corresponding contribution to the
mean magnetic pressure. Finally, $q_g$ refers to the anisotropic
contribution to the mean turbulent pressure owing to gravity.
The effective magnetic pressure
(the sum of turbulent and non-turbulent contributions
to the large-scale magnetic pressure)
is related to $q_p$ via
\begin{eqnarray}
\Peff = \frac{1}{2}(1-q_p)\beta^2,
\end{eqnarray}
where $\beta=\meanv{B}/\Beq$.

We compute $q_p$ by performing a reference simulation without an
imposed field to find $\Pi_{ij}^{\rm (f,0)}$ and a set of simulations
with a mean field to determine $\Pi_{ij}^{{\rm (f,}\meanv{B}{\rm
    )}}$ for a given field strength. Using Eq.~(\ref{equ:dPij})
in the $x$-direction
we find that it is sufficient to measure $\Delta\Pi_{xx}^{\rm (f)}$, from
which we obtain
\begin{eqnarray}
q_p = -2\mu_0 \Delta\Pi_{xx}^{\rm (f)}/\meanv{B}^2.
\end{eqnarray}
This expression agrees with that used in earlier work
\citep{LBKR14}.

Our measurements of the effective magnetic pressure $\Peff$ detected
negative values in the bulk of the convection zone, roughly consisting
of
80 \% of the deepest parts.
In the uppermost 20 \% of the domain $\Peff$ is always positive;
see a representative result in Fig.~\ref{fig:pPeff_prof_ms1024a2_rss}
from Run~C2v.
The effective
magnetic pressure in the middle regions of the layer between depths
$2.3$ and $7.0$~Mm for all the runs in Set~A are shown in
the top panel of Fig.~\ref{fig:pPeff_all}.
In the present convection setups, the equipartition field
strength is almost a constant throughout the layer due to which the
curves in Fig.~\ref{fig:pPeff_all} appear roughly as vertical lines
-- especially for weak imposed fields.
Taking data from the same depths in runs with different $B_0$ show a
trend which is very similar to that seen in forced turbulence with a
negative $\Peff$ for weak magnetic fields and positive $\Peff$
when the imposed field approaches equipartition, see the lower panel
of Fig.~\ref{fig:pPeff_all}.

The growth rate of NEMPI is proportional to the derivative of $\Peff$
with respect to the mean magnetic field strength:
\begin{eqnarray}
\lambda \propto \left( -2 \frac{d\Peff}{d \beta^2} \right)^{1/2},
\end{eqnarray}
see \cite{KeBKMR13} for an imposed horizontal field
and \cite{BGJKR14} for an imposed vertical one.
We find that in most of our simulations the derivative
of the effective magnetic pressure with respect to $\beta^2$
is positive (i.e., $d\Peff/d\beta^2 > 0$) almost everywhere in
the convection layer,
see a representative result from Run~C2v in the lower panel of
Fig.~\ref{fig:pPeff_prof_ms1024a2_rss}. In the runs with the strongest
imposed vertical fields $d\Peff/d\beta^2$ is negative in the
lower parts of the convection zone. In Runs~B3v and C3v this regime covers
roughly half of the depth of the layer.
The difference between the current simulations
and the density-stratified forced turbulence
models is that in our case the equipartition strength of the field is
almost constant in the bulk of the convection zone (see the lower
panel of Fig.~\ref{fig:purms}) whereas in the latter
$\Beq\propto \sqrt{\rho}$.
This leads to the situation that here $\beta$
varies relatively little in the bulk where $\Peff<0$,
and the derivative $d\Peff/d\beta^2$ has the wrong sign
for the excitation of NEMPI.
We have not tried to devise a situation where the
derivative $d\Peff/d\beta^2$ would be suitable for instability,
although this could perhaps be achieved by using imposed or
dynamo-generated fields that vary with height.

\begin{figure}[t]
\centering
\includegraphics[width=\columnwidth]{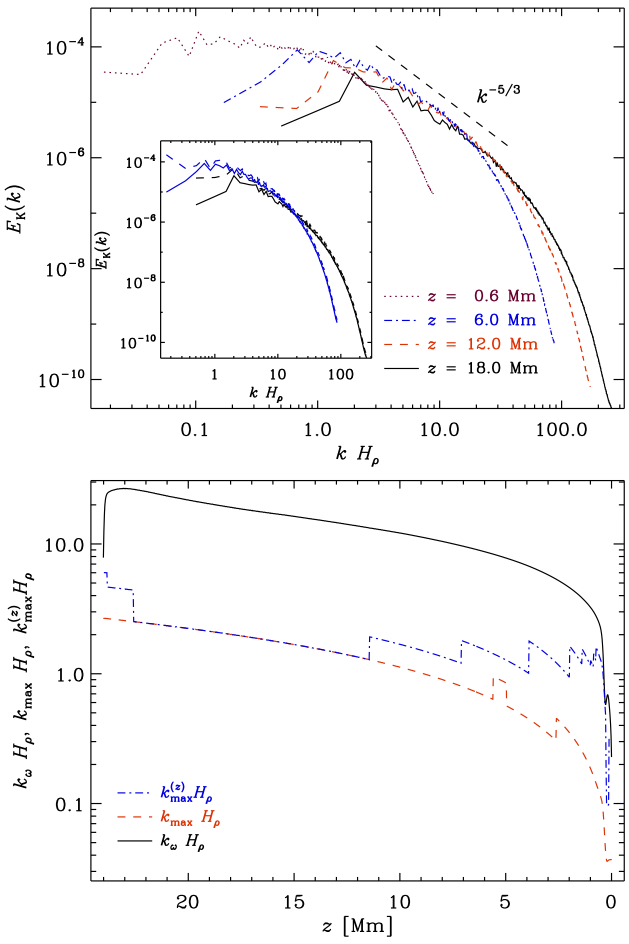}
\caption{Top panel: Power spectra of the fluctuating velocity from four
  horizontal planes as indicated
  in the legend in Run~C1v. The horizontal wavenumber is made
  non-dimensional by multiplying with the density scale height $\Hrho$ at the
  same depth. The dashed line shows the slope for Kolmogorov
  $k^{-5/3}$ scaling. The inset shows a comparison of power spectra of
  the full velocity field (dashed lines) and the fluctuating velocity
  from which the temporal average is removed (solid lines) from two
  depths. Bottom panel: wavenumbers corresponding Taylor microscale
  (black solid line), see Eq.~(\ref{equ:komega}), and the peaks of the
  fluctuating velocity power spectra (red dashed) and the fluctuating
  vertical velocity (blue dash-dotted) spectra as functions of depth
  and normalised by $\Hrho$.}
\label{fig:ppspectra}
\end{figure}

In addition to a negative derivative $d\Peff/d\beta^2$,
the scale separation ratio of turbulence needs to be sufficiently large for
the excitation of NEMPI.
DNS of forced turbulence \citep{BKKMR11,BKR13} show that, to excite NEMPI,
the scale separation ratio between the forcing scale and the size of the box
should be larger than 15.
Unlike the case of forced turbulence where
the forcing scale can be chosen as desired, the dominant scale of
turbulence in convection has to be estimated from the non-linear
outcome of the instability. This can be achieved by finding the peak of
the power spectrum of velocity. Convection is known to generate
large-scale circulations that are considered large-scale structures
rather than turbulence \citep[e.g.][]{EKRZ02,EKRZ06}. Thus, we first extract
the fluctuating part, $\uuu'$, by subtracting the average
velocity obtained by adding five snapshots separated by roughly half
a large-scale convective turnover time. We show power spectra
of the fluctuating velocity at four depths for Run~C1v in
Fig.~\ref{fig:ppspectra}. We also show a comparison with the spectra
from the full velocity field, showing that the power at large scales
is significantly reduced. We find that, near the surface and at a
depth of $6$~Mm, the spectra peak at the largest possible scale that
fits into the simulation domain. In the deeper layers, the peak is
found near $k H_\rho \approx 2$, which is of the same order of
magnitude as in \citep{KeBKMR13}. A similar estimate is found also for
the near-surface layers from the power spectra of the vertical
velocity, see the lower panel of Fig.~\ref{fig:ppspectra}. This is in
contradiction with the estimate obtained from the Taylor microscale,
i.e.\ $k_\omega H_\rho$, see Eq.~(\ref{equ:komega}), which is
typically an order of magnitude higher than $k_{\rm max}$
corresponding to the peak of the fluctuating velocity spectra.
In contrast to earlier lower resolution simulations
\citep[e.g.][]{CH06,KKB08}, we find a clear inertial range appearing
at intermediate scales in the deeper layers.

Previous work on NEMPI showed evidence of an intermediate phase
during which the magnetic field at large scales (characterizing
the large-scale structures) grows exponentially.
This was possible to see by isolating the large-scale magnetic field
through appropriate Fourier filtering.
By contrast, the total magnetic field, which includes the small-scale
magnetic field, grows linearly in time, which is expected when turbulence
acts on the applied magnetic field through tangling.
The exponential evolution of the large-scale field was taken
as evidence for the existence of a large-scale instability.
To check whether similar evidence can be produced here as well, we study
the early evolution of the largest scale Fourier components of the
vertical magnetic field near the surface of Run~C1v;
see Fig.~\ref{fig:pBB3_growth_ms1024a1_rss}.
However, it turns out that we do not find clear
evidence of exponential growth for any wavenumber.
The data is more consistent with linear growth suggesting that the
structure formation is related to tangling of the field by large-scale
convection. The lower panel of Fig.~\ref{fig:pBB3_growth_ms1024a1_rss}
shows the comparison of the two largest scale Fourier modes of $B_y$
in Run~A3h and a corresponding runs without backreaction to the
flow. In the latter case NEMPI cannot occur as the field is passive
and does not contribute to turbulent pressure. We find no significant
difference in the growth of the large-scale modes in these cases. This
suggests that even though we find a negative contribution to the
effective magnetic pressure in Run~A3h, NEMPI is not excited in the
simulation.
We conclude that the lack of clear exponential growth of the
structures in all of runs suggests that even though the sign of
$d\Peff/d\beta^2$ is favourable to NEMPI in some cases, the
instability is not excited.

In an earlier study, \cite{KKWM10} attribute the growth of magnetic
structures to vortical flows at the vertices of convection cells. They
also state that `usually the process starts at one of the strongest
vortices.' We note that in the simulations of \cite{KKWM10} the aspect
ratio of the box is close to unity. Comparing to our runs with
aspect ratio four we find that only a few large-scale convection cells
are present in the deep layers, see
Fig.~\ref{fig:puu_slices_ms1024a1}. This
suggests that most likely only a single large-scale convection cell
exists in the simulations of \cite{KKWM10}. This is not obvious from
the flows at the surface where several vortical downflows, which are
all connected to same large-scale downflow at deep layers, can be
identified. Thus, in their case a single downflow plume is likely
dominating the dynamics and concentrates the magnetic field, which is
consistent with the interpretation in terms of flux expulsion.

\section{Conclusions}

We demonstrate that stratified turbulent convection leads to
concentrations of magnetic field from an initially uniform field. The
area that these concentrations occupy in the volume is roughly
proportional to the imposed field strength. We also show that the
average size of the structures
increases with the box size when the imposed field strength is kept
constant. The strength of magnetic structures at large scales is
linearly proportional to the imposed field for horizontal fields. For
imposed vertical fields we find the same dependency for the smallest
domain size, whereas in larger domains the maximum approaches a
constant value. We also find a negative contribution to the effective
magnetic pressure -- in agreement with earlier studies
of turbulent convection
\citep{KBKMR12,KBKMR13}. However, the magnetic field in the
concentrations does not grow exponentially at any wavenumber, but is
consistent with linear growth. This indicates that
formation of magnetic concentrations is here not
associated with an instability such as NEMPI. We find that the
magnetic concentrations appear in regions where downflows associated with
large-scale, i.e.\ supergranular, convection occur. This process is
more commonly known as flux expulsion \citep{W66,GPW77,TWBP98}.
However, the role of turbulence in such flux expulsion is not yet clear.

The inability of the current simulations with an order of magnitude
greater density stratification than in our earlier studies to excite
NEMPI can be due to several reasons.
The excitation of NEMPI requires a negative sign of the derivative
of the effective magnetic pressure
with respect to the large-scale magnetic field.
In many cases in our simulations this derivative was positive, i.e.,
unfavourable for NEMPI.
In addition, it is possible that the
separation of scales between the system size and the turbulent scale
is insufficient (which in our simulations is only between $1$--$2$
when measured from the peak of the velocity power spectra, while in
forced turbulence the scale separation ratio of around 15 is needed
to observe NEMPI). Furthermore, convection in the current setup tends
to always develop also at the largest possible scale, which increases
as the domain size increases, and which dominates the generation of
magnetic concentrations. If this tendency carries over to the Sun, a
naive assumption is that giant cells of the order of 200~Mm should be
present and that they
would dominate the process of magnetic structure formation. Although
detection of giant cells in the Sun has been reported
\citep[e.g.][]{HUC13}, helioseismology appears to indicate a gaping
discrepancy between the Sun and current numerical simulations in that
the latter produce significantly too much power at large scales
\citep{HDS12}. Thus, at least circumstantial evidence suggests that a
new paradigm of convection could be needed. A possible candidate is
the concept of `entropy rain' \citep{Sp97,Br15} where only a thin top
layer, perhaps only a few Mm, of the convection zone is Schwarzschild
unstable and the rest of the layer is mixed by strong downflows
pummeling deep into the stably stratified interior. In such scenario
the largest scale excited by convection would be of the order of the
depth of the Schwarzschild unstable layer, and thus very much smaller
than in the current simulations where typically the whole domain is
unstable. This would eliminate giant cells and also
increase the scale separation drastically, perhaps enabling
NEMPI. However, devising numerical models capturing this idea is
challenging.

Another future step is to study formation of magnetic structures
in turbulent stratified convection from
the dynamo-generated field similarly to that for a
forced turbulence \citep{MBKR14,JBLKR14,JBKMR15}.

\begin{acknowledgements}
  The simulations were performed using the supercomputers hosted by
  CSC -- IT Center for Science Ltd.\ in Espoo, Finland, who are
  administered by the Finnish Ministry of Education. Special Grand
  Challenge allocation NEMPI12 is acknowledged. Financial support from
  the Academy of Finland grants No.\ 136189, 140970, 272786 (PJK) and
  272157 to the ReSoLVE Centre of Excellence (MJM), as well as the
  Swedish Research Council grants 621-2011-5076 and 2012-5797, and the
  Research Council of Norway under the FRINATEK grant 231444
  are acknowledged.
\end{acknowledgements}

\bibliographystyle{aa}
\bibliography{../bibtex/bib}

%\vspace{1cm} \noindent {\small \emph{$ $Id: paper.tex,v 1.230 2015/11/11 21:19:52 pkapyla Exp $ $}

\end{document}